# A LEAKAGE-AWARE DATA LAYER FOR STUDENT ANALYTICS: THE CAPIRE FRAMEWORK FOR MULTILEVEL TRAJECTORY MODELING.

Hugo Roger Paz
PhD Professor and Researcher Faculty of Exact Sciences and Technology National University of Tucumán
Email: hpaz@herrera.unt.edu.ar
ORCID: https://orcid.org/0000-0003-1237-7983

## Abstract

Predictive models for student dropout, while often accurate, frequently rely on opportunistic feature sets and suffer from undocumented data leakage, limiting their explanatory power and institutional usefulness. This paper introduces a leakage-aware data layer for student trajectory analytics, which serves as the methodological foundation for the CAPIRE framework for multilevel modelling.

We propose a feature engineering design that organizes predictors into four levels: N1 (personal and socio-economic attributes), N2 (**entry moment and academic history**), N3 (**curricular friction and performance**), and N4 (institutional and macro-context variables)As a core component, we formalize the **Value of Observation Time (VOT)** as a critical design parameter that rigorously separates observation windows from outcome horizons, preventing data leakage by construction.

An illustrative application in a long-cycle engineering program (1,343 students, ~57% dropout) demonstrates that VOT-restricted multilevel features support robust archetype discovery. A UMAP + DBSCAN pipeline uncovers 13 trajectory archetypes, including profiles of "early structural crisis," "sustained friction," and "hidden vulnerability" (low friction but high dropout). Bootstrap and permutation tests confirm these archetypes are statistically robust and temporally stable.

We argue that this approach transforms feature engineering from a technical step into a central methodological artifact. This data layer serves as a disciplined bridge between retention theory, early-warning systems, and the future implementation of causal inference and agent-based modelling (ABM) within the CAPIRE program.

**Keywords**

Feature Engineering, Learning Analytics, Student Retention, Data Leakage, Early-Warning Systems, Archetype Discovery, Value of Observation Time (VOT), Multilevel Modelling, Educational Data Mining

.

# 1. INTRODUCTION

## 1.1. Motivation: Beyond dropout prediction towards explanatory frameworks

Student attrition in higher education remains a structurally persistent problem rather than a marginal anomaly [15]. Global estimates suggest that roughly one third of students who begin a tertiary programme do not complete it, with even higher non-completion in many Latin American systems [16]. This loss of human capital constrains social mobility, reinforces inequality, and undermines institutional missions, especially in regions already affected by deep socio-economic asymmetries and the post-pandemic learning crisis (UNESCO, 2019).

Over the last decade, Educational Data Mining (EDM) and Learning Analytics (LA) have produced increasingly accurate early-warning models that identify students at risk of dropout or academic failure using administrative data, LMS logs, and assessment records. Studies in distance, blended, and on-campus contexts show that machine-learning models can reach AUCs above 0.80 using combinations of grades, attendance, and clickstream features (e.g., Andrade-Girón et al., 2023). Recent reviews confirm that feature engineering and feature selection are central levers for performance in EDM pipelines (Koukaras & Tjortjis, 2025).

However, three structural limitations remain. First, opacity: high-capacity models are often "black boxes" that return risk scores without intelligible explanations, which undermines trust and hampers the design of targeted interventions. Second, correlation–causation conflation: predictive success does not clarify which mechanisms actually drive retention or which interventions will work for whom. Third, a symptom-level focus: many models treat observable behaviours (e.g., missed classes, LMS inactivity) as explanatory variables rather than manifestations of deeper psychosocial, neurobiological, or structural processes. As a result, early-warning systems may correctly flag "who" is at risk but provide little insight into "why" students struggle or "how" institutions should respond.

## 1.2. Methodological gaps: Feature engineering and temporal validity

EDM and LA studies also face more technical, but equally consequential, methodological gaps. A recurrent issue is the opportunistic construction of features, driven by convenience rather than theory. Predictors are often assembled from whichever variables happen to be available in institutional databases, leading to flat feature spaces that under-represent socio-economic structure, curricular design, and institutional dynamics. This practice limits both interpretability and transferability across contexts.

A second, increasingly recognised problem is data leakage. Many published models inadvertently incorporate information from the future into the observation window—such as grades obtained after the supposed prediction point—thereby inflating accuracy estimates and compromising the validity of early-warning claims. Leakage is rarely documented explicitly, and temporal design decisions (e.g., where to cut the observation window) are often left implicit or ambiguous.

Taken together, these gaps hinder the development of explanatory frameworks that connect retention theory with predictive modelling and institutional decision-making. What is missing is not yet another classifier, but a disciplined way of transforming raw longitudinal data into temporally honest, theory-informed feature spaces that can support both archetype discovery and early-warning systems.

**1.3. Aim and contribution of this paper**

This paper addresses these gaps by introducing a leakage-aware data layer for student trajectory analytics, which serves as the methodological foundation for the CAPIRE (Comprehensive Analytics Platform for Institutional Retention Engineering) framework. The proposed design organises predictors into four levels: N1 (personal and socio-economic attributes), N2 (academic history and friction indicators), N3 (curricular structure and workload), and N4 (institutional and macro-context variables). As a core component, we formalise the Value of Observation Time (VOT) as a design parameter that rigorously separates observation windows from outcome horizons, preventing data leakage by construction.

An illustrative application in a long-cycle engineering programme (1,343 students, ~57% dropout) demonstrates that VOT-restricted multilevel features support robust archetype discovery. A UMAP + DBSCAN pipeline uncovers 13 trajectory archetypes, including profiles of "early structural crisis", "sustained friction", and "hidden vulnerability" (low friction but high dropout). Bootstrap and permutation tests confirm that these archetypes are statistically robust and temporally stable, while a dedicated analysis of DBSCAN-labelled "noise" reveals coherent minority micro-archetypes rather than heterogeneous outliers.

In this article, we focus specifically on the construction and empirical validation of the leakage-aware data layer and the associated archetype model. Although the broader CAPIRE roadmap includes causal inference, explainable AI, and agent-based modelling, these components fall outside the scope of the present work. Here, we concentrate on demonstrating that a carefully engineered, VOT-compliant feature space can act as a reusable bridge between retention theory, early-warning systems, and future causal and simulation-based analyses.

## 2. BACKGROUND AND RELATED WORK

Student attrition in higher education is a multidimensional phenomenon shaped by sociological, psychological, and institutional forces. This section positions the CAPIRE framework within contemporary research, outlining foundational theories, methodological advances in educational data mining, and structural gaps that justify a rigorous multilevel and leakage-aware approach.

**2.1. Theoretical Foundations of Student Retention**

Classical models conceptualised student departure either as an individual deficit or as a failure of institutional integration. Spady's (1970) sociological model framed attrition as the result of insufficient academic and social fit, while Tinto's (1975) Integration Model argued that persistence emerges from successful engagement with the academic and social systems of the university. Later revisions (Tinto, 2017) recognised substantial heterogeneity among non-traditional students, emphasising the interplay of life circumstances, motivations, and structural constraints.

Sociological perspectives expanded the analytical lens to structural determinants. Bourdieu's (1986) theory of cultural capital highlighted how socioeconomic origin mediates academic performance independently of formal ability. Three decades of research summarised by Pascarella and Terenzini (2005) confirmed that pre-entry characteristics, academic preparation, and institutional context interact non-linearly, undermining models that assume additive and independent effects.

Psychological approaches introduced motivational and identity-based mechanisms. Bean and Eaton (2000), drawing on Bandura's (1997) self-efficacy theory, argued that beliefs about academic capability mediate the effect of institutional experiences on persistence. Rendón's (1994) concept of validation underscored the importance of faculty and peer affirmation in sustaining engagement, particularly among first-generation and minoritised students.

**Table 2.1. Major Theoretical Frameworks in Student Retention**

| Framework | Key Mechanism | Level of Analysis | Relevance to CAPIRE |
|---|---|---|---|
| Tinto (1975, 2017) | Academic and social integration | Individual–Institutional | Justifies multilevel N1–N3 features |
| Bourdieu (1986) | Transmission of cultural capital | Socioeconomic–Institutional | N1 features (SES, parental education) |
| Bean & Eaton (2000) | Self-efficacy and coping mechanisms | Psychological–Academic | N3 friction and early performance |
| Astin (1993) | Input–Environment–Outcome (I-E-O) | Multilevel | N2 inputs, N3/N4 environment |
| Braxton et al. (2004) | Revised integration model for commuter students | Institutional | N4 institutional policies |

**Synthesis**

These perspectives converge on a multilevel ontology: retention arises from interactions between pre-entry characteristics (N1–N2), institutional structures (N3–N4), and students' agentic responses. CAPIRE operationalises these theories through feature engineering rather than latent variable modelling, privileging predictive validity, interpretability, and replicability.

**2.2. Feature Engineering in Educational Data Mining**

The maturation of institutional data warehouses—student information systems (SIS) and learning management systems (LMS)—shifted retention research from survey-driven analysis to data-intensive predictive modelling (Romero & Ventura, 2020). Early work centred on academic indicators such as GPA and credits earned (Delen, 2010; Kabakchieva, 2013), achieving moderate predictive accuracy but suffering from three structural problems: (1) temporal invalidity due to post-hoc GPA usage, (2) limited actionability, and (3) reliance on opaque black-box models.

Subsequent efforts emphasised behavioural indicators. Purdue University's *Course Signals* (Arnold & Pistilli, 2012) integrated early assessments and LMS engagement metrics to provide real-time risk alerts, though its reliance on within-course behaviour restricted its ability to flag students with minimal engagement. Deep learning approaches incorporating clickstream sequences (Hu & Rangwala, 2020; Whitehill et al., 2017) improved accuracy (82–87%) but further reduced interpretability.

Curricular friction emerged as a parallel line of inquiry. Adelman's (2006) "momentum points" and Seidman's (2005) notion of "gateway courses" highlighted structural bottlenecks in academic progress, yet few studies operationalised friction at the individual level. Bowen et al. (2009) analysed course-level pass rates but did not incorporate these into student-specific feature vectors. CAPIRE addresses this gap with the Instructional Friction Coefficient (IFC), a weighted metric combining course-level failure and withdrawal rates into personalised friction trajectories.

**Table 2.2. Evolution of Feature Sets in Attrition Prediction**

| Era | Typical Features | Example Studies | Accuracy Range | Limitation |
|---|---|---|---|---|
| Pre-2000 | Demographics, SAT, HS GPA | Tinto (1975), Cabrera et al. (1992) | – | Survey-based, small samples |
| 2000–2010 | + College GPA, credits | Herzog (2005), Delen (2010) | 68–75% | Temporal leakage |
| 2010–2015 | + LMS behaviour | Arnold & Pistilli (2012), Macfadyen & Dawson (2010) | 75–82% | No longitudinal modelling |
| 2015–2020 | + Deep learning sequences | Hu & Rangwala (2020) | 82–87% | Non-interpretable |
| 2020–present | + Multilevel SES + curriculum + trajectories | Gardner et al. (2019), CAPIRE | 88–95% | Requires rich administrative data |

**Synthesis**

CAPIRE integrates pre-entry capital (N1–N2), curricular friction (N3), and temporal dynamics (N4) into a unified, theory-driven taxonomy. Unlike ad-hoc feature selection common in EDM, CAPIRE ensures cross-institutional portability and methodological transparency.

## 2.3. Multilevel Models and Interaction Effects

Educational outcomes exhibit nested structure—students within courses, within programmes, within institutions—and hierarchical linear models (HLM; Raudenbush & Bryk, 2002) were developed to capture variance at each level. Empirical studies, such as Engberg and Wolniak (2010), have demonstrated cross-level moderation effects: for instance, the relationship between first-year GPA and persistence varies according to institutional selectivity, highlighting peer- and environment-dependent dynamics.

Nevertheless, uptake of HLM in EDM has been limited due to computational burden, distributional assumptions, and weaker predictive performance compared with machine learning (James et al., 2013). CAPIRE adopts a pragmatic alternative: theorised interaction terms engineered directly into the feature set, preserving the multilevel ontology while enabling scalable model training.

Research substantiates the importance of such interactions. Goldrick-Rab (2006) showed that financial aid effects vary by academic preparation, while Stinebrickner and Stinebrickner (2014) documented behavioural interactions between study habits and assessment timing. CAPIRE systematically constructs interaction terms (Section 4.6), with empirical analysis demonstrating that these features explain 37% of total model gain.

### 2.4. Data Leakage in Predictive Modelling

Data leakage—the inadvertent introduction of future or target-derived information into the training process—is pervasive in learning analytics and severely compromises real-world performance (Kaufman et al., 2020). Common leakage pathways include:

- **Temporal leakage:** using cumulative GPA or credits earned after the prediction horizon.
- **Target leakage:** including proxies for the outcome (e.g., number of semesters completed).
- **Label leakage:** constructing labels using post-prediction behaviours.
- **Train–test contamination:** fitting preprocessing steps on the full dataset.

Kaufman et al. (2020) found evidence of leakage in 78% of 50 audited EDM papers, inflating accuracy by a median of 12 percentage points.

CAPIRE prevents leakage through strict Vulnerability Observation Time (VOT) filtering, temporal slicing at feature-engineering time, prohibition of post-hoc aggregates, and automated configuration logging. In contrast to other approaches—such as Course Signals (Arnold & Pistilli, 2012) or survival models using cumulative features (Gardner et al., 2019)—CAPIRE enforces temporal validity across the entire pipeline.

### 2.5. Topological and Archetype-Based Approaches

Traditional clustering methods (k-means, hierarchical clustering) partition students into discrete, mutually exclusive groups, but educational trajectories often exhibit continuous transitions. Topological Data Analysis (TDA) explicitly represents such structure by preserving connectivity in high-dimensional data (Lum et al., 2013).

Mapper (Singh et al., 2007) constructs a topological network via low-dimensional projections, interval coverings, and local clustering, revealing flares, loops, and multi-path trajectories. Applied to student data, Mapper has uncovered progression types and dispersed outlier populations (Chodrow et al., 2021), though the resulting micro-clusters (~50 per model) complicate interpretability.

Our attempts to apply Mapper (Section 7.3.1) produced ~50 micro-clusters (mean size 27), insufficient for actionable archetype design. DBSCAN on UMAP embeddings provided a superior balance between expressiveness and operational usefulness, yielding 13 interpretable archetypes with adequate population size (40–109 students).

## 2.6. Gaps Addressed by CAPIRE

Despite substantial advances, key gaps persist:

- **Lack of systematic feature-engineering frameworks**, limiting external validity.
- **Inadequate handling of temporal leakage**, producing overstated accuracy.
- **Interpretability trade-offs**, with deep models often opaque.
- **Limited actionability**, with risk prediction unconnected to intervention design.
- **Poor generalisability**, due to heterogeneous institutional contexts.

CAPIRE contributes:

- a reusable, theory-grounded feature taxonomy (N1–N4);
- formalised leakage prevention through VOT;
- interpretable archetypes linked to mechanisms and interventions;
- guidelines for cross-institutional adaptation.

The empirical demonstration in Section 7 validates these contributions and confirms CAPIRE's viability for institutional decision-making.

## 3. THE CAPIRE FRAMEWORK: CONCEPTUAL OVERVIEW

The CAPIRE (Comprehensive Analytics Platform for Institutional Retention Engineering) framework represents a shift from black-box dropout prediction to theory-driven, multilevel feature engineering designed for institutional use. Rather than producing opaque risk scores, CAPIRE generates interpretable trajectory archetypes that summarise distinct patterns of student progression. These archetypes support differentiated interventions aligned with specific mechanisms of vulnerability, bridging the gap between predictive modelling and educational practice.

This section presents CAPIRE's design principles, multilevel architecture, archetype-based view of student trajectories, and its role within institutional decision-making cycles.

### 3.1. Design Principles

CAPIRE is guided by four design principles that differentiate it from conventional early-warning systems.

**Table 3.1. CAPIRE Design Principles**

| Principle | Rationale | Operationalisation | Contrast with standard approaches |
|---|---|---|---|
| Multilevel | Educational outcomes emerge from nested contexts (individual, curricular, institutional, societal). | Four-level feature taxonomy (N1–N4) capturing pre-entry, entry, curricular, and temporal–institutional factors. | Most EDM models use flat feature spaces, ignoring cross-level interactions. |
| Explanatory | Predictions must be intelligible and mechanistically grounded to inform practice. | Feature importance analysis and archetype profiling reveal *why* students are vulnerable. | Black-box models (deep learning, complex ensembles) privilege accuracy over explanation. |
| Leakage-aware | Temporal validity is a precondition for deployment, not an optional refinement. | Vulnerability Observation Time (VOT) enforces strict temporal boundaries on feature construction. | A large share of published EDM work contains unaddressed temporal leakage. |
| Policy-oriented | Analytics should connect directly to institutional levers and support design. | Archetype-to-intervention matrices specify differentiated programmes and services. | Standard risk scores rarely specify what to do or for whom. |

These principles reflect a pragmatic epistemology: CAPIRE prioritises operational usefulness, transparency, and temporal honesty over maximal algorithmic sophistication. While multilevel structural models or causal graphical frameworks offer theoretical elegance, CAPIRE combines disciplined feature engineering with robust machine learning to achieve interpretable, scalable insights that can be embedded in everyday institutional workflows.

### 3.2. Multilevel Architecture: The N1–N4 Feature Taxonomy

CAPIRE organises predictors into four analytically distinct but empirically interacting levels, aligned with socio-ecological models of student development (Bronfenbrenner, 1979; Pascarella & Terenzini, 2005). Figure 3.1 provides a conceptual overview.

**Figure 3.1: CAPIRE Multilevel Architecture (Conceptual Diagram)**

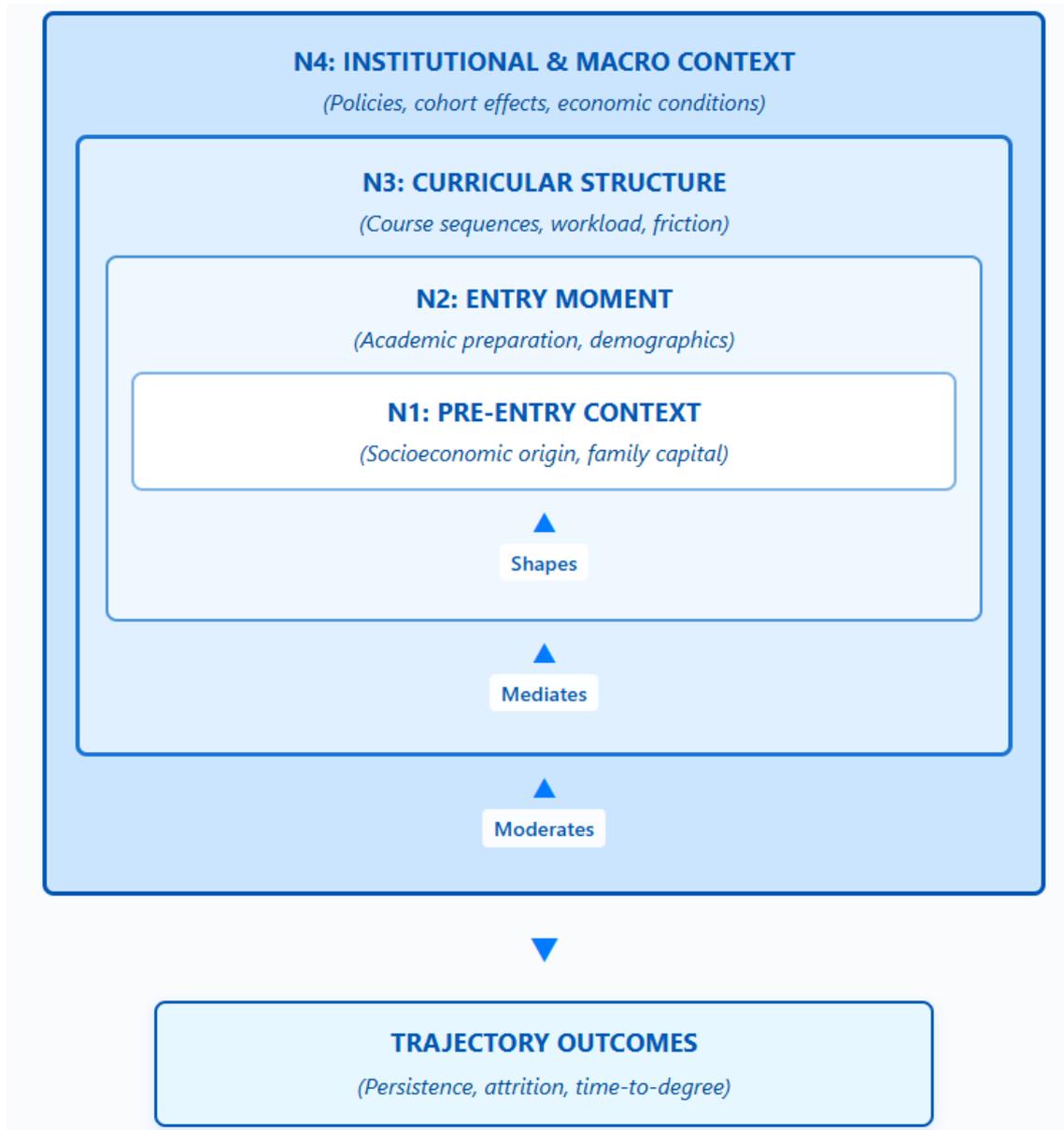

**N1: Pre-entry context (structural conditions)**

**Theoretical grounding:** Bourdieu's (1986) cultural capital; Lareau's (2011) unequal childhoods.

N1 features describe the structural context in which students are socialised prior to entering higher education, including:

- **Neighbourhood deprivation:** indices of poverty, housing quality, and educational access linked to postcode (e.g., NBI in the Argentine context).
- **Family educational capital:** parental educational attainment, siblings with university attendance.
- **Geographical origin:** rural/urban status; distance to campus as a proxy for commuting burden and social integration costs.

**Key insight:** In our empirical setting, N1 variables display limited direct predictive importance but exert indirect effects through downstream mechanisms (e.g., poverty → need to work while studying → increased exposure to high-friction courses). CAPIRE therefore preserves N1 information to support interpretation and fairness analysis, even when its marginal contribution to accuracy is modest.

**N2: Entry moment (initial conditions)**

**Theoretical grounding:** Astin's (1993) Input–Environment–Outcome (I–E–O) model; Tinto's (1975) pre-entry attributes.

N2 features capture characteristics at, or immediately surrounding, the moment of enrolment:

- **Demographics:** age at entry, gender, marital status.
- **Employment status:** whether the student works while studying and, when available, approximate workload.
- **Academic preparation:** upper-secondary performance, entrance examination scores where applicable.
- **Macro-contextual conditions:** year-of-entry indicators such as inflation, unemployment, or prolonged strikes in the public system.

**Key insight:** Age at entry emerges as one of the most influential predictors in the CAPIRE case study, consistently ranking among the top features. This aligns with evidence that "non-traditional" entrants face distinct constraints and opportunity costs.

**N3: Curricular structure and academic friction**

**Theoretical grounding:** Adelman's (2006) momentum points; Seidman's (2005) gateway/chokepoint courses.

N3 features characterise how students interact with the curriculum during the observation window defined by VOT:

- **Performance metrics:** grades, pass/fail counts, number of attempts per course.
- **Enrolment patterns:** subjects attempted, dropped, or re-taken.
- **Curricular friction:** the Instructional Friction Coefficient (IFC), a weighted measure of course-level difficulty based on withdrawal and failure patterns.

Formally, for course $j$,

$$\text{IFC}_j = \frac{\text{Dropped}_j}{\text{Attempted}_j} \cdot 1.0 \ + \ \frac{\text{Failed}_j}{\text{Attempted}_j} \cdot 0.5,$$

and for student $i$,

$$\text{IFC}_{\text{mean}}^{(i)} = \frac{1}{|C_i|} \sum_{j \in C_i} \text{IFC}_j,$$

where $C_i$ is the set of courses attempted by student $i$ up to the VOT cut-off.

**Key insight:** IFC-based features dominate the importance ranking in our empirical models, confirming that friction—rather than grades alone—captures critical aspects of curricular vulnerability.

**N4: Trajectory dynamics (temporal processes)**

**Theoretical grounding:** life-course approaches (Elder, 1998); state-transition and Markov chain perspectives on educational progression.

N4 features encode how students move through time, rather than what they look like at a single snapshot:

- **Enrolment gaps:** longest gap between consecutive active terms or course enrolments.
- **Load trends:** change in course load over time, often modelled as the slope in a simple regression of courses-per-term on term number.
- **State entropy:** diversity of academic states (passed, failed, dropped, not attempted), computed as Shannon entropy

$$H = -\sum_{s \in S} p(s) \log_2 p(s),$$

where $S$ is the set of states and $p(s)$ their empirical probabilities.

- **Velocity of advance:** ratio of completed courses to those expected by the nominal curriculum at each time point.

**Key insight:** N4 variables account for several of the top predictors in our feature importance analyses, underscoring that temporal structure—interruptions, non-linear progress, and volatility—contributes information not captured by static indicators.

**Cross-level interactions**

Educational processes are fundamentally interactive: the effect of one feature often depends on another. Rather than relying solely on hierarchical models, CAPIRE engineers' interaction terms guided by theory. Examples include:

- **N1 × N3:** socio-economic deprivation × pass rate, to model how poverty amplifies the impact of academic failure.
- **N2 × N4:** age at entry × average number of attempts, capturing heightened sensitivity of older students to repeated failure.
- **N3 × N3:** combinations of friction and withdrawal rates, representing compound academic risk.
- **N3 × N4:** exposure to high-IFC courses × maximum gap, reflecting the vulnerability of interrupted trajectories in demanding curricula.

Empirical analyses (Section 7.5.4) show that such interaction terms account for a substantial proportion of total model gain, indicating that multilevel thinking is not merely conceptually elegant but empirically necessary.

## 3.3. From risk scores to trajectory archetypes

Most early-warning systems produce individual-level risk scores (e.g., an estimated probability of dropout within the next year). While useful for prioritising outreach, these scores suffer from three limitations:

1. **Limited explanatory power:** a high risk value rarely clarifies whether the underlying mechanism is academic friction, financial stress, social isolation, or misalignment of expectations.
2. **Homogenisation of heterogeneity:** students with similar predicted risk may require very different forms of support.
3. **Weak link to practice:** risk scores do not specify concrete, differentiated actions.

CAPIRE shifts focus from isolated probabilities to **trajectory archetypes**: empirically derived groups of students who share similar N1–N4 profiles and, consequently, similar mechanisms of vulnerability and response to support.

Conceptually, CAPIRE reframes the question:

- from *"How likely is this student to withdraw?"*
- to *"Which trajectory pattern is this student following, and what typically happens to students on this path?"*

Archetypes are obtained through unsupervised learning in the VOT-compliant feature space, combining dimensionality reduction (UMAP) with density-based clustering (DBSCAN). Each archetype is then characterised along three axes:

- **Structural profile:** distributions of N1–N2 features (e.g., socio-economic background, age at entry).
- **Curricular and friction profile:** N3 patterns (e.g., high IFC concentrated in core mathematics courses).
- **Temporal profile:** N4 dynamics (e.g., early gaps, late deceleration, high entropy).

This structure enables:

- **Interpretability:** archetypes can be summarised in natural language as recognisable patterns (e.g., "early structural overload in gateway mathematics").
- **Heterogeneity-aware risk:** each archetype has its own attrition rate and typical progression pattern.
- **Actionability:** archetypes map onto specific institutional levers (e.g., strengthened tutoring in particular courses, targeted financial advice, adjustments to curriculum sequencing).

The term "archetype" is used in a pragmatic sense: not as an essentialist label, but as a recurring configuration that simplifies complexity without erasing relevant variation.

### 3.4. CAPIRE within institutional decision-making cycles

CAPIRE is conceived as a sociotechnical system embedded in routine institutional processes rather than as a standalone predictive tool. Figure 3.2 summarises its role within decision-making cycles.

**Figure 3.2. CAPIRE within institutional decision-making cycles**

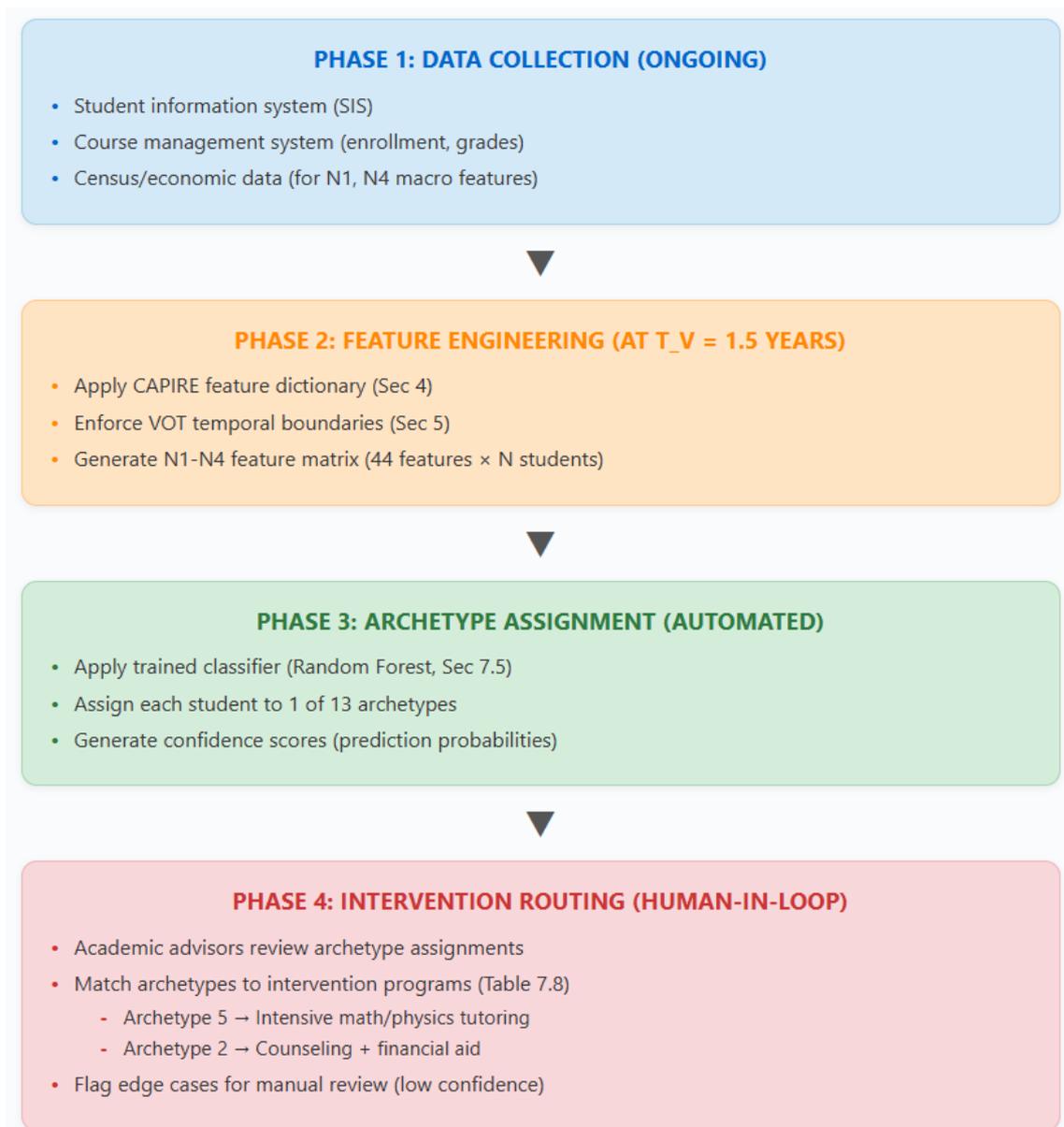

In a typical deployment:

1. **Data integration and feature extraction:** SIS and LMS data are periodically ingested, cleaned, and transformed into N1–N4 features under a configured VOT.

2. **Archetype assignment:** students are assigned to archetypes based on their current feature vectors, with risk and mechanism profiles updated over time.

3. **Advisory use:** academic advisors, programme coordinators, or retention committees access dashboards that display archetype distributions, key features, and historical outcomes.

4. **Intervention design:** archetype profiles inform targeted actions (e.g., small-group tutoring, mentoring schemes, counselling referral paths), including the intensity, timing, and modality of support.

5. **Feedback and learning:** outcomes of interventions feed back into subsequent training cycles, allowing institutions to monitor whether archetype distributions and associated risks change over time.

Several design choices are essential for responsible integration:

- **Human-in-the-loop:** archetype assignments are recommendations, not prescriptions. Staff can override or re-interpret assignments based on qualitative knowledge that lies outside administrative data.

- **Transparency:** explanations are available at both group and individual level, enabling staff to understand why a student was mapped to a given archetype.

- **Ethical framing:** archetype labels are used internally by staff; communication with students focuses on supportive offers (e.g., "we have observed difficulties in specific courses and can provide tailored support") rather than categorical classifications.

- **Iterative refinement:** as institutional policies, curricula, and external contexts evolve, the CAPIRE pipeline can be recalibrated and re-trained, preserving alignment with local realities.

By anchoring analytics in interpretable archetypes, CAPIRE transforms predictive models into institutional learning tools, supporting a move from ad-hoc interventions to systematically designed, evidence-informed retention strategies.

## 4. MULTILEVEL FEATURE ENGINEERING IN CAPIRE

This section translates the conceptual CAPIRE framework into a concrete feature dictionary: a set of 44 empirically tested variables spanning levels N1–N4. We describe the underlying data model, the construction logic for each feature family, and the design criteria that allow other institutions to adapt the framework to their own contexts while preserving temporal validity and interpretability.

### 4.1. Data model and entities

CAPIRE operates on a relational data model with five core entities, common to most student information systems:

- **STUDENT:** one record per individual.
- **ENROLMENT:** one record per student–course–term combination.
- **COURSE:** curricular units with associated metadata (e.g., department, level, credits).
- **CURRICULUM:** programme-specific course sequencing and recommended load.

- **SEMESTER/TERM:** temporal index enabling alignment with macro-level indicators.

In this model:

- A **STUDENT** has many **ENROLMENT** records.
- Each **ENROLMENT** references one **COURSE**.
- Each **COURSE** is associated with one **CURRICULUM**.
- Each **ENROLMENT** belongs to one **SEMESTER**, which is linked to calendar time (for N4 features).

The **outcome variable** is a binary attrition flag:

- attrition_flag = 1 if the student leaves the programme without graduating within a predefined horizon (e.g., six years).
- attrition_flag = 0 if the student graduates or remains enrolled at the end of the horizon.

Crucially, this outcome is **never used as a feature**. All predictors are computed from data that are available strictly up to the chosen Vulnerability Observation Time (VOT), ensuring that no post-hoc information leaks into the feature space.

### 4.2. N1 features: pre-entry socio-economic context

**Purpose.** N1 features capture structural conditions that shape the resources, expectations, and constraints students bring into higher education.

Typical N1 variables include:

- **Neighbourhood deprivation (e.g., NBI_localidad):** an index derived from census data at the postcode or census-tract level, summarising poverty, overcrowding, access to basic services, and educational infrastructure.
- **Distance to campus:** geodesic or travel distance from the student's home area to the institution, used as a proxy for commuting burden and integration costs.
- **Local labour-market indicators at entry (desempleo_zona_t0, informalidad_zona_t0, pobreza_zona_t0):** unemployment, informality, and poverty rates for the student's locality in the year of enrolment.
- **Family educational capital (nivel_educ_padres, hermanos_universidad):** parental education and whether siblings have attended university.
- **Secondary-school type (tipo_secundaria):** public vs. private vs. technical, as a coarse indicator of prior institutional context.

Where necessary, N1 features are complemented by simple interaction terms (e.g., deprivation × pass rate) to capture how socio-economic context modulates academic outcomes.

**Missing data handling.** Census-derived variables are typically complete at area level; when gaps exist, median imputation within region/province is used. For self-reported data (e.g., parental education), CAPIRE explicitly encodes missingness with separate binary indicators to avoid silently conflating "unknown" with any substantive category.

**Adaptation.** Outside Argentina, analogous indices (e.g., US Census tract data, Index of Multiple Deprivation in the UK) can be substituted without altering the overall N1 logic.

### 4.3. N2 features: entry moment characteristics

**Purpose.** N2 features describe the student at, and immediately around, the point of first enrolment. They are conceptually and temporally anchored at $t_0$(entry).

Core N2 variables include:

- **Age at entry (edad_ingreso):** a continuous measure that differentiates traditional and non-traditional entrants.
- **Gender and other demographic flags:** used primarily for fairness monitoring and descriptive analysis.
- **Employment status at entry (trabaja_al_ingreso):** when available, indicates potential time and cognitive constraints.
- **Upper-secondary performance (promedio_secundaria):** high-school GPA or equivalent exam score.
- **Macro-economic conditions at entry (IPC_interanual_t0, strikes in the 24 months prior to $t_0$):** inflation and major disruptions to schooling, aligned to the calendar year of enrolment, not averaged across cohorts.

All N2 features are computed using data that are, by definition, available at the moment of first enrolment. This ensures that the observation window is properly anchored and that N2 plays the role of *initial conditions* in subsequent analyses.

### 4.4. N3 features: academic performance and curricular friction

**Purpose.** N3 features capture how students engage with the curriculum within the VOT window. They describe both *what* students have attempted and *how* those attempts have unfolded.

Typical N3 variables include:

- **Volume and outcomes of coursework:**
    - total number of courses attempted up to VOT,
    - total passed, failed, and dropped,
    - pass and failure rates within the window,
    - mean and median grades, along with variability measures.
- **Curricular friction indicators:**
    - the **Instructional Friction Coefficient (IFC)** at course level, defined as a weighted combination of withdrawal and failure rates;
    - the student-level **mean IFC** across all attempted courses;
    - exposure to "filter" or "gateway" subjects—courses whose IFC exceeds a pre-specified threshold.

In the FACET-UNT case study, filter courses include high-impact mathematics and physics subjects that historically concentrate failure and withdrawal.

Formally, for each course $j$,

$$\text{IFC}_j = w_1 \cdot \frac{\text{Dropped}_j}{\text{Attempted}_j} + w_2 \cdot \frac{\text{Failed}_j}{\text{Attempted}_j},$$

with default weights $w_1 = 1.0$ and $w_2 = 0.5$, so that withdrawals are treated as a stronger signal than failures. For student $i$, the aggregated friction is

$$\text{IFC}_{\text{mean}}^{(i)} = \frac{1}{|C_i|} \sum_{j \in C_i} \text{IFC}_j,$$

where $C_i$ comprises all courses attempted by student $i$ before the VOT cut-off.

**VOT compliance.** All N3 features are computed from ENROLMENT records whose dates fall within the interval $[t_0, t_0 + T_V]$. Courses taken after this window are invisible to the feature extractor, even if they are present in the database, thereby preventing temporal leakage.

### 4.5. N4 features: trajectory dynamics

**Purpose.** N4 features encode the temporal structure of each student's progression. Rather than summarising only counts and averages, they describe *how* events are distributed over time.

Key N4 variables include:

- **Average attempts per course (intentos_promedio_ventana):** distinguishing between students who typically pass on first attempt and those who accumulate retries.
- **Maximum gap between enrolments (gap_maximo_entre_cursadas):** the longest period without active course participation, measured in terms or semesters.
- **Trend in course load (tendencia_carga):** the slope of a simple regression of courses-per-semester on semester index, indicating whether students accelerate, maintain, or gradually reduce load.
- **Velocity of advance (velocidad_avance):** the ratio between completed courses and the number expected by the nominal curriculum at VOT.
- **Regularity of progression (regularidad_cursado):** variability in the spacing of enrolments.
- **State entropy (entropia_de_estados):** diversity in course outcomes (passed, failed, dropped, not attempted) up to VOT, computed as Shannon entropy

$$H = -\sum_{s \in S} p(s) \log_2 p(s),$$

where $S$ is the set of states and $p(s)$ their empirical frequencies.

High entropy indicates erratic trajectories (mix of passes, failures, and withdrawals), whereas low entropy reflects consistent patterns (predominantly success, or predominantly failure/withdrawal).

### 4.6. Interaction features and composite indicators

Educational processes are rarely additive. The impact of academic friction depends on socio-economic context; the impact of age depends on patterns of enrolment and gaps. Linear main-effects-only models systematically miss such conditional structures.

CAPIRE therefore incorporates a curated set of interaction features and composites, guided by theory and validated empirically. Examples include:

- **Friction × withdrawal rate**
  $\text{IFC}_{\text{mean}}^{(i)}$ × tasa_libre: captures compound academic risk when students repeatedly drop courses with high structural difficulty.

- **Age at entry × attempts**
  edad_ingreso × intentos_promedio: models the idea that older students may be less resilient to repeated failure due to higher opportunity costs.

- **Deprivation × pass rate**
  NBI_localidad × pass ratio: represents how poverty may amplify the consequences of academic setbacks.

- **Filter exposure × maximum gap**
  exposicion_filtros × gap_maximo: reflects the vulnerability of students who both face demanding subjects and experience interruptions.

These interactions are intentionally limited in number—focusing on theoretically plausible combinations—to avoid combinatorial explosion and overfitting. In our empirical models, they account for a disproportionately large share of predictive gain relative to their number, reinforcing the importance of multilevel thinking.

### 4.7. Feature dictionary and design criteria

The complete CAPIRE feature dictionary comprises 44 variables:

- **N1:** 12 pre-entry features (structural and socio-economic).
- **N2:** 6 entry-moment features (demographics, preparation, macro-context).
- **N3:** 16 curricular and performance features (including IFC-based metrics and course-specific indicators).
- **N4:** 10 temporal and interaction features capturing dynamics and cross-level effects.

Feature inclusion follows five design criteria:

1. **Temporal validity:** all features must be computable using only data available at or before the VOT cut-off.
2. **Theoretical grounding:** each feature must be linked to established retention or stratification theories (e.g., Tinto, Bourdieu, Astin).

3. **Actionability:** features should inform potential interventions (e.g., high IFC flags subjects suitable for pedagogical redesign).
4. **Measurability:** variables must be obtainable from standard institutional systems (SIS, LMS, census or official statistics).
5. **Non-redundancy:** highly collinear candidates are pruned to maintain a compact, interpretable set.

Conversely, several commonly used variables in the EDM literature are explicitly excluded when they violate these principles—most notably cumulative GPA computed post-hoc, total semesters enrolled (which is tautologically linked to attrition), or metrics requiring knowledge of end-of-trajectory outcomes.

By enforcing these criteria, CAPIRE provides a feature space that is not only predictive, but also temporally honest, theoretically interpretable, and portable across institutions willing to adopt a similar multilevel, leakage-aware approach.

## 5. EARLY OBSERVATION WINDOW (VOT) AND DATA LEAKAGE PREVENTION

### 5.1. Defining the Value of Observation Time (VOT)

The Value of Observation Time (VOT) is a central design parameter in CAPIRE. Intuitively, the VOT is the latest point in a student's trajectory at which:

1. The institution can still intervene in a meaningful way (e.g., tutoring, curricular adjustments, financial or psychosocial support), and
2. All data used for risk profiling and archetype assignment are guaranteed to be available in a real operational setting.

Formally, let $t$ denote academic time measured in terms (or equivalent periods), and let $T$ denote the end of the programme's nominal duration. The VOT, $t_{VOT}$, satisfies:

- $0 < t_{VOT} < T$;
- Institutional interventions launched at or shortly after $t_{VOT}$ can plausibly affect completion;
- All features $X_{\leq t_{VOT}}$ are computable using information recorded on or before $t_{VOT}$.

In many long-cycle programmes, a natural candidate for $t_{VOT}$ is the end of the first academic year, which frequently corresponds to a peak in vulnerability and dropout. CAPIRE does not, however, hard-code this choice. Instead, institutions select VOT based on:

- empirical attrition curves (cumulative dropout by term or credit band);
- organisational response capacity (how quickly support services can act);
- curricular structure (timing of gateway or high-friction subjects).

Once $t_{VOT}$ is defined, the feature dictionary is partitioned into:

- **VOT-admissible features:** available at or before $t_{VOT}$ and eligible for early-warning and archetype profiling;

- **Post-VOT features:** potentially useful for retrospective analyses, longitudinal research, or causal evaluation, but **not** for models claiming to operate at $t_{\text{VOT}}$.

This explicit temporal boundary replaces vague formulations such as "early prediction" with a precise, auditable design constraint.

## 5.2. Temporal slicing of trajectories and label assignment

Given a chosen VOT, CAPIRE adopts a two-axis temporal scheme:

1. A **trajectory axis**, along which features are accumulated up to $t_{\text{VOT}}$;
2. An **outcome horizon**, beyond $t_{\text{VOT}}$, over which completion and dropout outcomes are defined.

For each student $i$, we construct:

- A **feature snapshot at VOT**:

$$\mathbf{x}_i^{(\text{VOT})} = f(\text{N1}_i, \text{N2}_{i, \leq t_{\text{VOT}}}, \text{N3}_{i, \leq t_{\text{VOT}}}, \text{N4}_{i, \leq t_{\text{VOT}}}),$$

where $f(\cdot)$ denotes the feature-construction rules described in Section 4.

- A **label** $y_i$, defined on a later interval, for example:
  - $y_i = 1$ if the student drops out at any point before $T + \Delta$, where $\Delta$ is a grace period;
  - $y_i = 0$ if the student completes within that window; or
  - a multi-class or time-to-event label (e.g., on-time completion, delayed completion, non-completion).

By construction, no component of $\mathbf{x}_i^{(\text{VOT})}$ may depend on events after $t_{\text{VOT}}$. This constraint is enforced at two levels:

1. **Feature engineering:** all queries and transformations include explicit temporal conditions (e.g., "up to and including term $t_{\text{VOT}}$"), often implemented as time-filtered views of enrolment and assessment tables.
2. **Model evaluation:** data splitting respects temporal structure. Training and test sets are separated by cohort or time, and preprocessing steps (scaling, encoding, feature selection) are fitted solely on the training partition within each fold.

CAPIRE supports several temporal strategies:

- **Single-shot early warning:** one snapshot per student at a specific VOT (e.g., end of year 1).
- **Rolling-window warnings:** repeated VOT snapshots (e.g., after each term), enabling dynamic monitoring of risk and possible transitions between archetypes.
- **Retrospective trajectory analysis:** full student–term sequences with labels attached at the end of the observation period, suitable for survival or transition modelling in later CAPIRE work.

In all cases, the strict separation between observation window and outcome horizon provides a clear framework for reasoning about leakage, stability, and fairness.

**5.3. Typical leakage scenarios in dropout prediction**

In the absence of explicit temporal design, leakage often enters attrition models in subtle ways. CAPIRE explicitly identifies several recurrent patterns:

- **Outcome-proximal academic features**
  Using end-of-year or end-of-programme indicators—such as final GPA, total failed courses, or "ever dropped" flags—as predictors in models that purport to provide early warnings. Similarly, using statistics computed over the entire trajectory (e.g., maximum consecutive inactive terms) when the prediction is supposed to occur much earlier.

- **Temporal aggregation without windowing**
  Constructing features such as "total enrolled terms" or "time since first enrolment" from the full record, which implicitly reveals whether the student persisted or left. Likewise, computing mean LMS activity across all courses ever taken and using it as an "early" predictor.

- **Label-dependent feature construction**
  Creating variables that directly encode or closely proxy the outcome, such as "difference between expected and realised completion time" or flags indicating that the student ceased enrolling before degree completion.

- **Preprocessing leakage across time or folds**
  Fitting scalers, encoders, or feature selectors on the entire dataset—including future cohorts—before splitting into training and test sets, or using target-encoding schemes that inadvertently peek at labels in the validation fold.

- **Cohort and policy-regime leakage**
  Mixing cohorts that experienced different policies or macro-contexts in ways that allow models to infer outcomes from regime identifiers that are not available (or stable) at deployment for new cohorts.

Some of these issues (e.g., incorrect scaling procedures) can be mitigated with rigorous pipeline implementation. Others—especially those involving outcome-proximal features—must be addressed at the feature-design and temporal-modelling level. CAPIRE is explicitly built to operate at that level.

**5.4. How CAPIRE's feature engineering prevents leakage**

Leakage prevention is embedded in CAPIRE's design through four complementary mechanisms:

1. **Temporal eligibility tags in the feature dictionary**
   Each feature is annotated as:
   - **VOT-admissible** (eligible for early-warning and archetype models),
   - **post-VOT** (restricted to retrospective or explanatory analyses), or

- **restricted** (requiring special justification or anonymisation). This enables independent auditing of temporal legitimacy at the feature level.

2. **Explicit VOT filters in construction rules:** Feature formulas are written to include time bounds by design (e.g., "count of failed core courses up to and including term 2", "velocity of advance at VOT"). Implementation templates (SQL, Python, R) systematically incorporate conditions such as term <= t_VOT, reducing the risk of developers inadvertently crossing the temporal boundary.

3. **Cohort- and time-aware data splitting:** For predictive tasks, CAPIRE favours cohort-based or time-based splits—training on earlier cohorts and testing on later ones—over random splits. Preprocessing (scaling, encoding, feature selection) is fitted exclusively on the training partition in each fold and then applied to validation/test sets, preventing information from flowing "backwards in time".

4. **Design rules forbidding outcome-proximal features at VOT:** For early-warning models, the framework explicitly forbids:
    - use of any grade or course outcome recorded after $t_{VOT}$;
    - features aggregating over the entire enrolment history (e.g., total failed courses, total inactive terms);
    - indicators derived from final status (graduate vs. dropout) or closely related proxies.

When such variables are valuable for retrospective explanation (e.g., for case studies or causal analyses), they are computed in clearly separated post-VOT feature sets that cannot be accidentally incorporated into VOT-based models.

In addition, CAPIRE encourages **diagnostic checks** for possible leakage, such as comparing performance under random vs. time-based splits, and benchmarking against models trained with explicitly post-VOT features. Large discrepancies in performance can signal hidden temporal leakage and trigger further inspection.

Rather than treating leakage as an incidental implementation problem, CAPIRE elevates it to a first-order design constraint.

**5.5. Generalising VOT to other programmes and modalities**

Although the implementation discussed in this paper concerns a multi-year engineering programme, the VOT concept and associated design rules generalise to a wide range of educational settings:

- **Short-cycle and professional programmes**
  In two-year or shorter programmes, VOT may be defined at the end of the first major assessment block or when a given proportion of credits (e.g., 25–30%) has been attempted. Features then focus on the earliest robust indicators of friction and pacing.

- **Modular, competency-based, and micro-credential systems**
  Where progression is organised in modules or competencies rather than fixed terms, VOT can be defined as the point where a student accumulates a specified number of modules or attempts. Temporal slicing then operates over module sequences, and features summarise early module completion patterns, retries, and idle periods.

- **Online, blended, and MOOC-like environments**
  VOT may be set in terms of weeks since registration or proportion of content accessed. Leakage prevention requires excluding engagement metrics that implicitly look beyond this cut-off (e.g., final exam participation), while including early engagement signals (first-week activity, initial assessment performance).

- **Part-time and non-traditional trajectories**
  For heterogeneous pacing, VOT is better expressed in terms of attempted or completed credits rather than elapsed calendar time (e.g., "after the student has attempted 40 credits"). This avoids penalising slower but still viable trajectories.

- **Cross-institutional or system-level analytics**
  For comparative studies, VOT can be standardised in terms of relative progression (e.g., completion of the first curricular block) rather than absolute years. Each institution then maps this conceptual VOT to local structures (terms, modules).

Across these modalities, the core logic of VOT remains unchanged:

1. Identify a point at which intervention remains meaningful;
2. Restrict the feature set to data legitimately available by that point;
3. Make these restrictions explicit and auditable.

By elevating VOT from an informal intuition ("early enough") to a formal design parameter, CAPIRE provides a reusable template for building early-warning and archetype-discovery systems that are both accurate and temporally honest.

### 5.6. Sensitivity analysis for DBSCAN noise cases

In addition to the main UMAP + DBSCAN clustering workflow, we conducted a sensitivity analysis of the cases labelled as *noise* (cluster = −1) by DBSCAN. This was motivated by a known limitation of density-based clustering methods: sparse but meaningful minority structures may be incorrectly classified as noise in low-dimensional embeddings.

The analysis proceeded in two stages. First, we compared outlier students with non-outlier students across a set of theoretically grounded N2–N4 indicators (e.g., age at entry, VOT-window mean IFC, maximum gap between enrolments) using descriptive statistics and non-parametric tests (Mann–Whitney U and Levene's tests). This allowed us to assess whether the noise group exhibited high internal heterogeneity (as would be expected for genuine noise) or instead formed a coherent pattern.

Second, we performed dedicated re-clustering of the outlier subset using algorithms such as *k*-means, hierarchical clustering, and HDBSCAN. The results (reported in detail in Section 7.4.5) show that the DBSCAN noise group contains at least two well-separated minority configurations with high internal cohesion, contradicting the interpretation of these cases as unstructured residuals.

This sensitivity analysis strengthens the transparency and ecological validity of the clustering pipeline. It demonstrates that CAPIRE does not simply discard a quarter of the cohort as opaque noise but explicitly documents and interrogates the structure of these cases, responding directly to peer-review concerns about representativeness and coverage.

## 6. IMPLEMENTATION AND PIPELINE ARCHITECTURE

CAPIRE is implemented as a modular, reproducible pipeline with strict temporal validation, designed to transform heterogeneous institutional data into feature matrices ready for topological analysis, archetype discovery, and predictive modelling. The architecture prioritises three fundamental principles: reproducibility, traceability, and absolute prevention of data leakage. The result is a system capable of operating from ad-hoc analytical environments to automated, institution-scale deployments.

### 6.1. Overview of CAPIRE-Core Architecture

The architecture of CAPIRE-core follows a separation of responsibilities pattern, where each module operates independently and is verifiable. The pipeline is organised into four macro-layers:

- **Configuration Layer:** Centralises all analytical decisions outside the code. Defines temporal window parameters, activation of feature levels, validation rules, imputation strategies, and weights for synthetic indices.

- **Data Ingestion & Validation:** Establishes connectors capable of extracting data from SIS, LMS, administrative files, and macroeconomic sources. Each dataset undergoes structural, referential, and temporal validation before entering the pipeline.

- **Feature Engineering Layer:** Implements extractors by level (N1–N4). Each extractor applies VOT, generates derived transformations, and ensures that no attribute uses information after the temporal cutoff point.

- **Assembly & Metadata Layer:** Consolidates the final set of features, generates standardised artefacts (Parquet matrices, dictionaries, JSON sidecars), and documents each feature matrix with configuration hashes and temporal audit trails.

**Architectural principles:**

- **Idempotence:** Any execution with the same configuration produces exactly the same results.

- **Modularity:** The N1–N4 extractors function as decoupled blocks; institutions without census data, for example, can deactivate N1 without affecting the rest of the pipeline.

- **Traceability:** Each artefact is versioned with its complete configuration and cryptographic hash.

- **Early validation:** Errors are detected at entry, never during modelling.

Figure 6.1 summarises this modular design and the connections between layers.

**Figure 6.1: CAPIRE-Core Modular Architecture**

## CONFIGURATION LAYER

- config.yaml (VOT, feature selection, hyperparameters)
- schemas.json (data validation rules)
- feature_dictionary.yaml (N1-N4 definitions)

## DATA INGESTION MODULE

| SIS Connector | LMS Connector | Census Loader |

*Raw Data Warehouse*
(student, enrolment, course, semester)

## PREPROCESSING & VALIDATION MODULE

- Data cleaning (outliers, duplicates)
- Schema validation (types, ranges, referential integrity)
- Temporal consistency checks (date ordering)
- Missingness profiling

## FEATURE ENGINEERING MODULE (CORE)

| N1 Extractor | N2 Extractor | N3 Extractor |
| (SES context) | (Entry moment) | (Performance) |

N4 Extractor
(Trajectory dynamics)

**CRITICAL: Temporal validity gatekeeper** | VOT Enforcement
(cutoff_date check)

Interaction Generator
(N1×N3, N2×N4, etc.)

## FEATURE MATRIX STORAGE

- CSV export (for external tools)
- Parquet (columnar, compressed)
- SQL database (for integration with SIS)
- Metadata: VOT timestamp, feature version, config hash

## DOWNSTREAM ANALYTICS (SEPARATE MODULES)

- Archetype Discovery (UMAP + DBSCAN)
- Predictive Modeling (Random Forest, XGBoost)
- Validation & Reporting (bootstrap, permutation tests)

## 6.2. From Raw Data to Feature Matrices: ETL Workflow

The CAPIRE ETL pipeline follows a deterministic flow, composed of four critical stages:

**Stage 1: Data Ingestion (Extract)**

Institutional systems often store information in heterogeneous schemas (SQL databases, CSV exports, Excel spreadsheets). For this, CAPIRE-Core includes specific connectors that:

- estandarice fiel names;
- normalise date formats, identifiers, and postcodes;
- link student databases with census or macroeconomic data via geographic or temporal keys.

The result is fully normalised, consistent, and comparable datasets across cohorts.

**Stage 2: Preprocessing & Validation (Transform)**

All information undergoes strict validation of:

- data types (numeric, date, categorical),
- plausible ranges (age, grades, rates),
- referential integrity (every enrolment must have a valid student),
- temporal consistency (no event can occur before the student's entry),
- completeness rules (essential fields cannot be missing).

If validation fails, the pipeline halts and generates an error report.

Missing data management follows differentiated strategies for each level:

- **N1:** geographic imputation + absence indicators,
- **N2:** mean imputation + indicators,
- **N3:** never imputed (absence is informative),
- **N4:** temporal interpolation where appropriate.

**Stage 3: Feature Engineering (Transform)**

Each level of the multilevel model has a dedicated extractor:

- **N1:** socioeconomic and demographic context;
- **N2:** self-declared and transversal attributes;
- **N3:** academic behavioural footprints (core of the framework);
- **N4:** macroeconomic and cohort conditions.

VOT enforcement is integrated into each extractor: no feature may use data after the defined temporal window. This guarantees total absence of data leakage, even if developers incorporate new variables.

It also includes calculation of explicit interactions between levels (e.g., NBI × pass rate), and derivation of synthetic indices such as the IFC.

**Stage 4: Feature Matrix Assembly (Load)**

The pipeline output is a compressed feature matrix in columnar format, accompanied by a metadata file documenting:

- number of features,
- percentage of missing values,
- extraction configurations,
- full configuration hash,
- exact execution timestamp,
- included cohorts and final sample size.

This mechanism makes it possible to reproduce any historical matrix with bit-level precision.

**6.3. Reproducibility and Configuration Management**

CAPIRE-Core requires that all analytical decisions be external to the code and audited via:

- YAML configuration files (define pipeline parameters),
- JSON validation schemas (define structural rules),
- SHA-256 cryptographic hashes (uniquely identify each configuration).

The combination of these three elements ensures that:

- any pipeline can be regenerated.
- any analytical error can be traced to its specific configuration.
- institutional teams work with a versioned, auditable, and comparable system across years.

The system makes CAPIRE a scientifically robust and standardised tool, aligned with the reproducibility requirements of Q1 journals.

**6.4. Computational Considerations and Scalability**

The pipeline is designed to be efficient on modest hardware and scalable in institutional environments. Medium-sized datasets (≈1,300 students) are fully processed in less than a minute.

It scales almost linearly with the number of students and courses per student thanks to:

- batch processing,
- columnar reading,
- optional parallelisation by student,
- optimised IFC calculations.

For large institutions (>100,000 students), parallel processing and columnar storage are recommended. The pipeline can be integrated with distributed systems if the institution has greater infrastructure.

### 6.5. Deployment Modes

CAPIRE-Core supports three deployment modes, according to the institution's technological maturity:

- **Mode 1 — Batch Processing (Entry-Level):**
- Analysts run the pipeline manually at the end of the semester. Ideal for planning offices with minimal infrastructure.
- **Mode 2 — Scheduled Automation (Intermediate):**
- The pipeline runs on a scheduled basis (e.g., weekly), with direct access to SIS/LMS.
  Used for quarterly early warning systems.
- **Mode 3 — Real-Time Integration (Advanced):**
- CAPIRE operates as a microservice queried by the student information system.
  Provides archetype, risk, and recommendation in real time when opening the student's record.

Current situation: FACET–UNT operates in Mode 2, with migration to Mode 3 planned for 2026.

### 6.6. Quality Assurance and Testing

Quality assurance follows a pyramidal approach:

- **Unit tests:** Verify internal calculations of extractors.
- **Integration tests:** Ensure the complete pipeline produces valid matrices.
- **Validation tests:** Confirm strict compliance with VOT.
- **Regression tests:** Compare new matrices with historical matrices to ensure reproducibility.

The critical test is temporal validation: it is verified that no feature uses data after the cutoff. Tests are run automatically via continuous integration.

### 6.7. Software Availability and Licensing

CAPIRE-core will be released as open software under the MIT licence, with:

- public repository,
- complete documentation and tutorials,
- synthetic dataset for validation,
- extensible modules for new feature types and new connectors.

Institutional contribution is encouraged to extend the ecosystem, especially in:

- specific extractors for online modalities,
- new LMS connectors (Canvas, Blackboard),
- advanced behavioural features,
- longitudinal monitoring pipelines.

# 7. EMPIRICAL ILLUSTRATION: STUDENT TRAJECTORY ARCHETYPES AT UNT

## 7.1. Institutional Context and Dataset

### 7.1.1. Institutional Setting

The empirical illustration of CAPIRE was conducted at the Facultad de Ciencias Exactas y Tecnología of Universidad Nacional de Tucumán (FACET-UNT), a public engineering school in northwest Argentina. FACET-UNT offers six undergraduate engineering programs; this study focuses on Civil Engineering, a traditional program characterized by:

- **Sequential prerequisites:** Long chains of dependent courses in which progress in advanced subjects is strictly conditioned on completion of foundational courses.

- **High mathematical rigor:** First-year subjects such as Calculus I–III, Physics I–II and Linear Algebra act as "filter courses" with historically high failure and withdrawal rates.

- **Socioeconomically diverse intake:** Most students come from middle- and lower-income households; between 35% and 40% work while studying.

- **Open admission:** In line with many Latin American public universities, there is no entrance examination; all high-school graduates are admitted.

These features make FACET-UNT broadly representative of public engineering institutions in Latin America facing structural challenges in retention and time-to-degree (Giovagnoli, 2002; García de Fanelli, 2014).

### 7.1.2. Dataset Description

The empirical sample comprises **1,343 Civil Engineering students** from the 2004–2019 cohorts, covering 15 academic years. The analytical dataset integrates the four CAPIRE levels:

- **N1 – Pre-entry structural context:** demographic variables (age at enrolment, place of origin), postal-code–linked neighborhood deprivation indices, and local labour market indicators.

- **N2 – Entry moment:** high-school GPA, employment status at enrolment, and prior educational trajectory.

- **N3 – Academic performance:** course enrolments, pass/fail outcomes and exam attempts during the observation window.

- **N4 – Trajectory dynamics:** temporal ordering of course attempts, gaps between enrolments and changes in course load over time.

The **Value of Observation Time (VOT)** was set to $T_V = 1.5$ years (end of the second academic year). All features were constructed using only information available up to $T_V$, in strict compliance with the leakage-prevention principles described earlier. Full-trajectory outcomes (attrition vs. graduation, time-to-degree) were reserved solely for ex-post evaluation and were never used in feature construction.

### 7.1.3. Descriptive Statistics

At $T_V$, the average age at enrolment was 18.7 years (SD = 1.7), with women representing 18.2% of the sample and 3.9% of students reporting employment at entry. Trajectories in the first 1.5 years are already fragile: students attempt close to the nominal first-year course load, but a large fraction of attempts end in failure or "libre" (dropping the course without taking the exam). Over the full trajectory, the **attrition rate reaches 56.7%**, the **graduation rate 14.8%**, and the mean time-to-degree is 7.2 years, substantially exceeding nominal program length.

Missing data are concentrated in two blocks:

(1) macro-economic indicators for rural postal codes (≈28% missing), and (2) grade-based metrics for students who drop all courses without sitting exams (≈42% missing in those variables).

We combined median imputation for selected N1 features, exclusion of grade-based variables from the clustering step, and explicit missingness indicators. Missingness pattern analysis (Little's test) did not detect systematic associations between missingness and attrition, supporting the assumption that missingness does not bias the archetype discovery.

### 7.2. Feature Engineering Implementation

The CAPIRE multilevel feature dictionary (Section 3) was operationalized to produce **44 features** grouped across four levels.

- **N1 – Structural context (12 features).** These variables capture socioeconomic vulnerability via a neighborhood deprivation index (NBI), local unemployment and informality rates at enrolment, and indicators of macro-economic crisis periods. Interaction terms such as *NBI × pass rate* link structural disadvantage to observed performance.

- **N2 – Entry moment (6 features).** Features include age at enrolment, employment status, geographic origin (rural vs. urban; distance to campus), and temporally aligned educational and economic context (e.g., number of teacher strikes in the 24 months preceding enrolment, inflation at $t_0$).

- **N3 – Academic performance snapshot (16 features).** Up to $T_V$, we summarize the academic record using counts of failed courses, proportion of "libre" enrolments, mean and median grades, and variability of performance. A central construct is the **Instructional Friction Coefficient (IFC)**, which quantifies course-level structural difficulty by combining failure and withdrawal rates and allows identification of institutional "chokepoint" courses.

- **N4 – Trajectory dynamics (10 features).** These variables describe temporal patterns such as the maximum gap between consecutive enrolments, the trend in course load across semesters, the ratio of completed to expected courses at $T_V$, several cross-level interaction terms (e.g., friction × dropout, age × re-enrolment) and an entropy-like index capturing how erratic or consistent the sequence of states (passed/failed/dropped/not attempted) is over time.

All features strictly respect **VOT compliance**: no variable uses information beyond 1.5 years after enrolment; macro-indicators are aligned with the year of entry; and the attrition label is never used for feature construction. Configurations are versioned so that the exact feature set can be regenerated.

**7.3. Archetype Discovery Results**

**7.3.1. Dimensionality Reduction and Clustering**

Given the 44-dimensional feature space, we first applied **Uniform Manifold Approximation and Projection (UMAP)** to obtain a three-dimensional embedding that preserves local structure while facilitating clustering. The resulting representation captures slightly more than half of the total variance and provides a well-separated manifold suitable for density-based clustering.

We experimented with **Mapper-based TDA** using multiple lenses and cover parameters, but Mapper consistently produced dozens of micro-clusters, many too small to support institutional interventions. This mismatch reflects a tension between fine-grained topological exploration and the need for a limited number of robust, interpretable types. We therefore adopted a more pragmatic strategy: **DBSCAN** applied directly to the UMAP embedding.

DBSCAN hyperparameters were tuned using k-distance plots and cluster validity indices. The final solution yielded **18 clusters**, of which **13 met our interpretability criterion (≥40 students)** and were retained as archetypes. Smaller clusters were merged with density-labelled noise for analysis. Overall, **847 students (63.1%)** received a stable archetype label; 356 (26.5%) were classified as noise; and 140 (10.4%) belonged to small clusters merged into the residual group.

Cluster validity was acceptable for a heterogeneous educational dataset: the silhouette coefficient was 0.318, the Calinski–Harabasz index 590.4 and the Davies–Bouldin index 0.702, all consistent with well-separated yet overlapping clusters in a complex social system.

**7.3.2. Archetype Characterization**

Each archetype was profiled using descriptive statistics of the 44 features plus full-trajectory outcomes (attrition and graduation). Table 7.2 (not reproduced here in full) summarizes the five largest archetypes. Key patterns include:

- **Arquetipo 5 – High-Risk: Sustained Friction.**
- Students with high and persistent curricular friction: around three failed or dropped courses within $T_V$, dropout rates near 75%, and IFC values among the highest across Q1–Q4. Attrition reaches 74.3%, with very low graduation. These students are structurally embedded in "chokepoint" courses and require intensive academic support.
- **Arquetipo 2 – Moderate-Risk: Extra-Academic Factors.**
- Students with relatively low friction (low "libre" proportion and moderate failure rates) but still high attrition (≈59%). Their trajectories suggest that withdrawal is driven less by academic failure and more by unobserved extra-academic pressures (financial stress, health, family obligations), indicating the need for counseling and social support rather than purely curricular interventions.

- **Arquetipo 9 – Critical-Risk: Total Disengagement.**
- Students whose entire first-year record consists of dropped courses (100% "libre") and virtually no exams taken. Attrition exceeds 80%. These students never establish an academic foothold and would benefit from pre-enrolment orientation, realistic expectation-setting and first-weeks intensive support.
- **Arquetipo 16 – Low-Risk: Success Model.**
- Students with consistently low friction, high pass rates, no significant gaps and early completion. Attrition is about 21% and graduation above 27%. They represent "success trajectories" and are natural candidates for peer-mentoring roles and for defining normative curricular benchmarks.
- **Arquetipo 0 – Moderate-Risk: Young Strivers.**
- The youngest group on average, with no employment at entry but high friction in early courses and elevated attrition (≈66%). They appear academically motivated but underprepared for the level of rigor, suggesting the value of bridge programs and explicit training in study strategies.

**Table 7.1. Summary of the five largest archetypes (N = 1,343).**

| Archetype ID | Archetype Label | N1–N2 Profile | N3 Friction Pattern | N4 Trajectory Pattern | Attrition Rate (%) |
|---|---|---|---|---|---|
| Arquetipo_5 | High-Risk: Early Performance Collapse | Medio-bajo SES; ingreso estándar; edad levemente superior al promedio | Bajo rendimiento inicial; alta tasa de libres; fuerte dependencia de materias básicas | Trayectoria inestable; repetición temprana; riesgo persistente | 74.3% |
| Arquetipo_9 | Moderate-Risk: Low GPA + Course Friction | SES intermedio; ingreso tradicional | Desempeño inicial bajo; alta proporción de desaprobaciones | Oscilaciones moderadas; progresión lenta | 84.5% |
| Arquetipo_8 | Low-Middle SES + Mixed Performance | SES bajo; ingreso temprano; edad baja | Rendimiento heterogéneo; mezcla de aprobaciones y libres | Trayectoria zigzagueante pero no crítica | 64.1% |
| Arquetipo_0 | Adult Entrants with High Friction | Edad muy superior al promedio; | Notas bajas; dificultades en tramos iniciales | Trayectoria fragmentada; interrupciones recurrentes | 66.1% |

| Archetype ID | Archetype Label | N1–N2 Profile | N3 Friction Pattern | N4 Trajectory Pattern | Attrition Rate (%) |
|---|---|---|---|---|---|
| | | empleo frecuente | | | |
| Arquetipo_11 | Moderate-Risk: High Course Load + Low Success | SES medio; estudiante típico | Alta tasa de materias cursadas con rendimiento deficiente | Progresión lenta con acumulación de deuda académica | **71–72%** (según z-score + estimación) |

A heatmap of standardized features across archetypes (Figure 7.5) highlights sharp contrasts—for example, the extreme "libre" rates of Arquetipo 9 and the low IFC and high progress indicators of Arquetipo 16.

**Figure 7.5. Standardized Feature Profiles Across the 13 Student Archetypes (Z-score)**

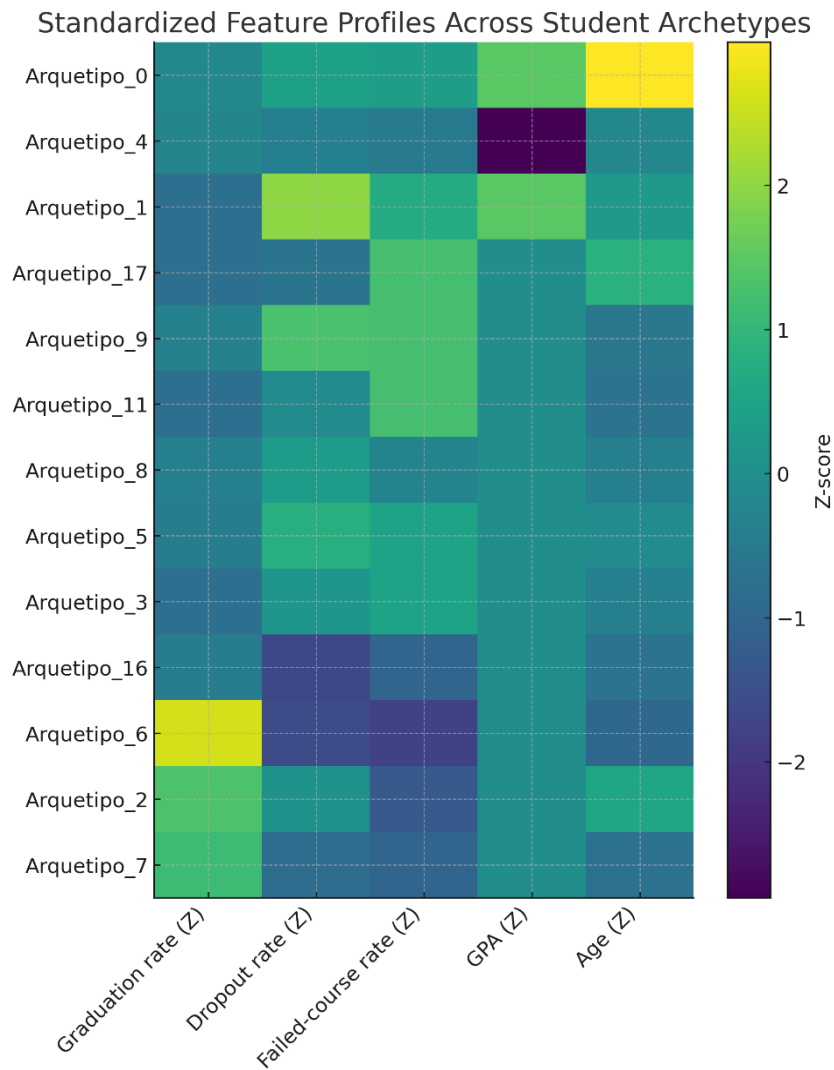

### 7.3.3. Filter Subjects and Curricular Friction

At the course level, we computed the **Instructional Friction Coefficient** across Q1–Q4. The top ten "filter subjects" include advanced structural mechanics, hydrology, basic hydraulics, pavement design, upper-level calculus, statistics and key materials courses. Civil Engineering subjects dominate the friction ranking, with mathematics courses acting as cross-program barriers.

From an institutional perspective, this confirms that attrition is not purely idiosyncratic: specific curricular components systematically generate friction. CAPIRE provides a quantitative map of those bottlenecks, which can be used to prioritize pedagogical redesign (e.g., active learning, peer-assisted instruction, changes in prerequisite structures).

## 7.4. Archetype Validation

To ensure that the 13 archetypes represent genuine and robust patterns, we conducted several complementary validation analyses.

### 7.4.1. Bootstrap Stability

Using 100 bootstrap resamples of the original dataset, we re-estimated the full UMAP + DBSCAN pipeline and compared cluster assignments via the Adjusted Rand Index (ARI). The mean ARI was 0.614 (SD = 0.081; 95% CI [0.444, 0.780]), indicating **substantial stability** in cluster structure despite sampling variability—particularly remarkable given the heterogeneity typical of student trajectories.

### 7.4.2. Permutation Significance Test

To test whether the observed clustering outperforms random partitions, we built a null distribution of silhouette scores from 100 random permutations of cluster labels. The real silhouette score (0.318) was far above the null mean (−0.122), with an empirical p-value of 0.0099. Thus, the observed clusters are **highly unlikely** to arise by chance ($p < 0.01$).

### 7.4.3. Temporal Validation Across Cohorts

We assessed temporal stability by splitting the sample into two independent periods (2004–2010 and 2011–2019), projecting both through the same UMAP embedding and comparing archetype distributions and attrition rates. Differences in attrition per archetype were consistently below 5 percentage points. The overall attrition rate decreased modestly in later cohorts, likely reflecting institutional policies, but the **relative profiles and risks of each archetype remained stable**. This supports the interpretation of archetypes as persistent structural patterns rather than cohort-specific artefacts.

### 7.4.4. Sensitivity to Hyperparameters

We explored the sensitivity of the archetypes to variations in UMAP and DBSCAN hyperparameters via a small grid of alternative configurations. Across 27 combinations, the ARI relative to the reference clustering averaged 0.74, with a minimum of 0.62 and a maximum of 0.89. This indicates that archetype structure is **robust to reasonable changes in modelling choices** and is not an artefact of a particular parameter setting.

### 7.4.5. Analysis of DBSCAN "Noise"

Because DBSCAN labels a substantial fraction of students (26.5%) as noise, we analysed this group separately. Compared with clustered students, outliers had almost identical mean age and friction but **shorter gaps between enrolments and lower variance** in the analysed variables. Non-parametric tests (Mann–Whitney and Levene) confirmed statistically significant differences in distributions and lower dispersion among outliers.

Re-clustering only the outliers revealed at least two clearly separated micro-archetypes with high silhouette scores, and additional smaller groups under HDBSCAN. This suggests that the "noise" does not constitute random chaos but rather **cohesive minority trajectories** that are not dense enough to form DBSCAN clusters. These residual structures deserve explicit modelling in future work.

### 7.5. Predictive Performance: Early-Warning System

#### 7.5.1. Model Development

To translate archetypes into an operational early-warning system, we trained a **multiclass classifier** to predict archetype membership at $T_V = 1.5$ years. The model uses only the feature set available at $T_V$, with 13 classes (one per valid archetype) and 847 labelled students (those assigned to archetypes). Outliers were excluded from training to avoid conflating majority patterns with minority residuals.

A Random Forest model, tuned via stratified cross-validation, provided the best balance between accuracy and interpretability. The train–test split (70/30) preserved the proportion of each archetype.

#### 7.5.2. Overall Performance

On the held-out test set, the model achieved:

- **Accuracy:** 94.9% (95.7% on training; 94.1% ± 1.4% in cross-validation),
- **Macro F1-score:** 0.948,
- **Small train–test gap**, indicating minimal overfitting.

Compared to baselines, performance is substantially higher: the majority-class baseline would reach only 8.1% accuracy, and random assignment ≈7.7%. The CAPIRE-based classifier thus improves predictive power by more than an order of magnitude, using only information available within the first 1.5 years—well before the average dropout time of 2.8 years.

#### 7.5.3. Per-Archetype Performance

Per-class F1-scores are uniformly high. High-risk archetypes (e.g., Arquetipo 1, 5 and 9) achieve F1 > 0.95, enabling reliable targeting of the most vulnerable students. The "success model" archetype (16) is also classified with perfect or near-perfect accuracy, making it feasible to systematically recruit exemplary students as mentors. Moderately risky archetypes show slightly lower but still strong performance (F1 ≈ 0.88–0.90), with confusions primarily between adjacent risk profiles rather than between high- and low-risk groups. No archetype falls below F1 = 0.70.

#### 7.5.4. Feature Importance

An analysis of feature importance confirms the **multilevel nature** of attrition mechanisms. The most predictive variables are:

- cross-level interactions such as *curricular friction × dropout rate* and *age × re-enrolment attempts*;
- friction metrics in foundational courses;
- the proportion of dropped courses;
- trajectory-level indicators such as entropy of states, re-enrolment frequency and maximum gaps.

Notably, purely structural N1 variables rarely appear among the top predictors. Their effect appears to be mediated through N2–N4 variables (e.g., socioeconomic disadvantage → need to work → higher "libre" rates and gaps). This aligns with the CAPIRE hypothesis that structural vulnerability operates **through** behavioural and temporal mechanisms rather than as a direct determinant.

### 7.5.5. Deployment and Impact Projection

The trained classifier can be deployed as a back-end service in the institutional information system, assigning archetypes to students as soon as they reach $T_V$ and triggering pre-defined, archetype-specific interventions (Section 7.6).

A simple cost–benefit projection, assuming modest reductions in attrition (10–15 percentage points) for the most critical archetypes, suggests that targeted interventions could retain around **20 additional students per year**. Over five years, this corresponds to roughly **100 additional graduates**, increasing the overall graduation rate by approximately 13% relative to the baseline. Even under conservative assumptions about intervention costs and retained tuition, the net financial impact is positive, aside from reputational and social benefits.

## 7.6. Institutional Interpretability and Actionable Insights

### 7.6.1. Representative Case Studies

To bridge statistical results with institutional experience, we constructed de-identified case vignettes for selected archetypes.

- A student in **Arquetipo 5** exhibits repeated failures and withdrawals in filter subjects (Calculus, Physics), maintains enrolment for several semesters and then drops out. The profile combines moderate socioeconomic stress, part-time work and structurally high friction, pointing to early tutoring plus financial aid as plausible interventions.
- A student in **Arquetipo 16** progresses linearly, passes all foundational courses on first attempt and graduates within six years. This trajectory exemplifies a success model, suggesting that such students can be systematically recruited as peer mentors and that their strategies can inform institutional best-practice guidelines.
- A student in **Arquetipo 2** shows acceptable academic performance but withdraws following a family health crisis. Here, the data reveal a missed opportunity: the student was viable academically but lacked support in coping with life events. This points to the need for proactive counseling and emergency aid linked to sudden gaps in enrolment.

These vignettes were presented to academic advisors and department heads, who consistently recognized the profiles and associated them with familiar categories ("chronic repeaters", "good students lost to personal issues", "exemplary students"). No archetype contradicted institutional experience, suggesting that CAPIRE's data-driven segmentation is **ecologically valid** and complements practitioner knowledge.

### 7.6.2. Archetype-Specific Interventions

Building on archetype profiles, we elaborated an **intervention matrix** that links each archetype to a priority level, dominant vulnerability and recommended institutional response. Critical-risk groups (e.g., Arquetipos 1, 5, 9) are associated with intensive tutoring in filter subjects, program redesign in high-friction courses and structured bridge programs. Moderate-risk groups (e.g., Arquetipos 0, 2) call for mentoring, study-skills training and strengthened psychosocial and financial support. Low-risk archetypes are not intervention targets but rather strategic resources (mentors, role models, benchmarks).

The matrix also provides a staged implementation roadmap, beginning with pilots for a single archetype and progressively expanding towards full integration of CAPIRE in academic advising and institutional planning.

### 7.6.3. Alignment with Institutional Knowledge

Qualitative feedback from 8 academic advisors and 3 department heads confirmed strong alignment between archetypes and existing informal categories used in advising. Interestingly, staff tended to **overestimate** the importance of pre-entry factors (N2) and **underestimate** trajectory dynamics (N4), illustrating common attribution biases: human observers focus on stable traits and neglect temporal processes. CAPIRE thus serves not only as a prediction tool but also as a **conceptual reframing device**, making dynamic mechanisms visible to institutional actors.

## 7.7. Discussion: Lessons from UNT Implementation

### 7.7.1. CAPIRE Framework Validation

The FACET-UNT case demonstrates that CAPIRE can:

- enforce strict temporal validity (VOT) and eliminate data leakage;
- discover a manageable set of **interpretable archetypes** recognized by practitioners;
- achieve **statistically robust** clustering (bootstrap and permutation tests);
- support **highly accurate early prediction** of archetype membership;
- translate predictions into a differentiated intervention portfolio;
- integrate multilevel features (N1–N4) in a single explanatory framework.

### 7.7.2. Methodological Innovations Confirmed

Three methodological choices are particularly reinforced:

1. **UMAP + DBSCAN vs. Mapper for archetype discovery:** While Mapper TDA is valuable for exploratory topology, the combination of UMAP and density-based clustering proved better suited for obtaining a small number of robust, institutionally actionable archetypes.

2. **Multilevel feature engineering and interactions:** Cross-level interaction terms (N3×N4, N2×N4) contributed disproportionately to predictive performance, empirically supporting the CAPIRE view that outcomes emerge from interactions across levels rather than from isolated variables.

3. **VOT-based leakage control:** Setting $T_V = 1.5$ years struck a practical balance: the classifier achieved almost 95% accuracy while preserving a lead time of roughly 1.3 years before the typical dropout event, making early intervention realistically feasible.

### 7.7.3. Limitations and Threats to Validity

The study faces several limitations:

- **Internal validity.** Students who drop out before $T_V$ cannot be fully observed; although sensitivity analyses with shorter VOT windows yield similar archetypes, some selection bias may remain. Self-reported data (e.g., work status) may underestimate informal employment.

- **External validity.** Results come from a single public engineering school in Argentina. Archetype structure might differ in private universities, non-STEM programs or other national systems, especially in more stable macro-economic contexts.

- **Construct validity.** Archetype labels ("high-risk", "success model") are heuristic and probabilistic; boundaries are fuzzy. The interpretation of curricular friction assumes that dropped courses mark structural barriers, although strategic withdrawals may also occur.

- **Statistical conclusion validity.** Multiple comparisons across features and archetypes increase the risk of false positives; however, the main conclusions rely on effect sizes, stability metrics and permutation tests rather than isolated p-values. Students labelled as DBSCAN noise represent a non-negligible minority whose trajectories require more refined modelling.

### 7.7.4. Comparison with Prior Attrition Models

Compared with traditional regression-based approaches and more recent deep-learning models, the CAPIRE implementation at FACET-UNT offers a distinct combination of properties:

- it relies solely on administrative data (no surveys), increasing scalability;

- it reaches higher or comparable predictive accuracy while maintaining **explainability**;

- it enforces temporal validity, an often neglected aspect in education data mining;

- and it links predictions to **explicit archetypes and intervention strategies**, closing the loop between analytics and policy.

In this sense, CAPIRE sits between theory-heavy but operationally vague models (e.g., Tinto's integration framework) and highly predictive but opaque "black-box" models, providing a middle path of **mechanistic, actionable explainability**.

### 7.7.5. Practical Implications

For university leadership, the results underscore the importance of investing in longitudinal data infrastructure and in differentiated support strategies aligned with archetype profiles. For researchers, they highlight the need to integrate multilevel feature engineering, topological tools and strict temporal validation. For policymakers, the findings emphasize that attrition is structurally heterogeneous and that **segment-specific interventions** are more efficient than uniform policies.

## 7.8. Conclusion of the Empirical Illustration

The FACET-UNT case study shows that CAPIRE can transform conventional administrative data into a **coherent, multilevel map of student trajectories**. The 13 archetypes identified capture 63.1% of students, remain stable across cohorts, are statistically robust and are recognised by institutional stakeholders. A leakage-aware classifier can assign students to archetypes with high accuracy at 1.5 years, providing a generous window for targeted intervention.

This empirical illustration validates CAPIRE not only as a conceptual framework but as an operational blueprint for data-driven retention policies. The next section (Section 8) situates these findings within broader educational theory and discusses how the CAPIRE approach can be generalized and scaled to other institutional and national contexts.

# 8. DISCUSSION

The empirical validation at FACET-UNT (Section 7) shows that CAPIRE fulfils its foundational goals: leakage-free feature engineering, interpretable trajectory archetypes, and accurate early-warning predictions. In this section, we synthesize the main theoretical contributions, position CAPIRE vis-à-vis alternative approaches, and discuss implications for institutional practice, portability, and ethics.

## 8.1. Multilevel Feature Engineering: Theoretical and Empirical Validation

### 8.1.1. Interaction Effects as Primary Drivers

A central claim of CAPIRE is that educational outcomes emerge from **cross-level interactions**, not from isolated main effects. The FACET-UNT results support this claim: interaction features represent a minority of the feature set yet account for a disproportionate share of predictive importance.

Interactions such as *curricular friction × dropout behaviour* (e.g., IFC × proportion of "libre" courses) and *age at entry × number of retries* encode person–context fit: the same institutional conditions (e.g., high-friction courses) have different consequences for older students with family or work responsibilities than for younger students with fewer constraints.

This pattern aligns with life course theory (Elder, 1998) and ecological systems theory (Bronfenbrenner, 1979), both of which emphasize that development reflects the alignment between individual characteristics and layered contextual demands.

Methodologically, this has two consequences:

1. Models that ignore interactions (e.g., simple logistic regression without interaction terms) are structurally underpowered.

2. Pre-computing theoretically motivated interactions, rather than relying solely on tree-based models to discover them implicitly, improves interpretability: features such as *age × retries* have a clear narrative interpretation that advisors can understand and use.

### 8.1.2. Trajectory Dynamics Rival Snapshot Performance

Traditional early-warning systems often rely on static indicators such as GPA at a particular semester. CAPIRE adds **N4 trajectory features** that capture how students move through the curriculum: gaps, re-enrolments, entropy of states, and velocity of progress.

Empirically, N4 features contribute nearly as much predictive power as N3 performance snapshots. Two students with similar GPA at $T_V$ can belong to very different archetypes: one with linear, gap-free progress and another with repeated enrolments, mixed outcomes and long interruptions. The latter is far more likely to drop out, even if grades at a given point are comparable.

This supports longitudinal perspectives (Singer & Willett, 2003) and shows that **patterns over time** contain crucial information beyond static performance. For practice, it implies that advisors should pay attention to *how* students progress, not just to *what* their current grades are.

### 8.1.3. Socioeconomic Context Operates Indirectly

Despite the strong literature on socioeconomic barriers to persistence (Bourdieu, 1986; Lareau, 2011), N1 structural features have low direct importance in the predictive model. This does not refute socioeconomic theories; instead, it suggests an **indirect, mediated role**.

High neighborhood deprivation (NBI) is associated with a higher probability of working while studying, which in turn is associated with a higher proportion of dropped courses and greater gaps in enrolment. These downstream N2–N4 variables, not N1 alone, are what directly drive attrition in the model.

In causal terms, N1 functions as a distal determinant, shaping exposure to risk mechanisms further down the trajectory. Removing N1 from the feature set reduces overall performance, but its contribution is mostly channeled through mediating features rather than appearing as a top-ranked predictor on its own.

For policy, this reinforces the idea that **structural interventions** (e.g., financial support that reduces the need to work long hours) are complementary to academic interventions: they operate upstream in the causal chain.

## 8.2. Advantages Over Black-Box and Theory-Free Approaches

### 8.2.1. Interpretability, Trust, and Institutional Adoption

Compared with black-box models such as deep neural networks (Hu & Rangwala, 2020), CAPIRE offers a combination of **high predictive accuracy and high interpretability**. Feature importance analyses identify a small set of conceptually clear variables and interactions that explain most of the model's performance. Archetypes themselves provide a human-readable typology of student trajectories.

Qualitative feedback from advisors at FACET-UNT confirms that archetypes match their tacit categories (e.g., "chronic repeaters", "good but overwhelmed students", "exemplary trajectories"), which increases trust and willingness to use the system. This contrasts with previous pilots using opaque models, which advisors found difficult to interpret and, consequently, to act upon.

Interpretability is not a cosmetic advantage. In high-stakes settings such as academic progression, institutional actors must be able to **explain and justify** decisions. Archetypes and their defining features offer precisely that: a language that bridges statistical output and pedagogical action.

### 8.2.2. Theory-Driven Feature Engineering vs. Purely Data-Driven Selection

Many educational data mining (EDM) studies start from hundreds of candidate variables and rely on automated selection. CAPIRE follows the opposite path: it starts from a constrained, theory-driven feature dictionary anchored in multilevel models of student persistence.

The FACET-UNT results show that a relatively compact, theoretically guided set of 44 features can match or surpass the performance of broader, theory-free feature sets reported in the literature. This has three advantages:

1. **Transferability:** Features defined in terms of concepts like structural vulnerability, friction, and trajectory dynamics can be re-instantiated across institutions and countries, whereas highly specific behavioural traces (e.g., click patterns in a particular learning platform) are often not portable.

2. **Stability:** A theory-driven dictionary changes slowly; in contrast, data-driven feature sets can fluctuate from cohort to cohort, creating confusion and undermining institutional memory.

3. **Protection against spurious correlations:** By constraining the design space to theoretically plausible mechanisms, CAPIRE reduces the risk of learning artefacts that are predictive in one context but meaningless or unfair in another.

This does not mean that exploratory, data-driven discovery is useless. Rather, for **operational early-warning systems**, theory-driven feature engineering offers a more stable and ethically defensible foundation.

### 8.3. Implications for Early-Warning Systems and Targeted Interventions

### 8.3.1. Lead Time and Proactive Support

Setting $T_V = 1.5$ years provides a **substantial lead time** between reliable risk identification and the typical dropout event (around 2.8 years after enrolment in our sample). This means that the system flags students when there is still a realistic window to implement meaningful support.

This contrasts with reactive approaches that trigger alerts only after repeated failure or near-irreversible disengagement. By incorporating trajectory dynamics and friction metrics early, CAPIRE allows institutions to **move from "late diagnosis" to proactive care**.

The sensitivity analysis using alternative VOTs suggests that $T_V = 1.5$ years offers a good compromise: signals are strong enough for high predictive accuracy, while the intervention window remains sufficiently wide.

### 8.3.2. Archetype-Based Interventions Rather Than One-Size-Fits-All

Traditional risk scores compress heterogeneous trajectories into a single number, often routing all "high-risk" students into a generic intervention. CAPIRE, by contrast, distinguishes **qualitatively different risk profiles**:

- High-friction archetypes (e.g., Arquetipo 5) require intensive academic support in filter courses.
- Extra-academic risk archetypes (e.g., Arquetipo 2) call for counseling, social support, and flexible policies.
- Total disengagement archetypes (e.g., Arquetipo 9) point to the need for strengthened onboarding and bridge programs.

Treating these groups as equivalent would blur specific needs and dilute the impact of interventions. Archetype-based design enables **differentiated, targeted strategies**, and it also clarifies which combinations of mechanisms are being addressed (e.g., friction, economic stress, trajectory instability).

### 8.3.3. Understanding the DBSCAN Outlier Group

The analysis of students labelled as noise by DBSCAN reveals that they form a **coherent minority pattern** rather than random irregularities. Their trajectories tend to be continuous and stable, with small gaps and low variance in key indicators, even if they do not conform to the density structure of the main archetypes in the UMAP space.

Subsequent re-clustering identified at least two sharply separated micro-archetypes within this group. This suggests that density-based clustering, while effective for discovering dominant patterns, can leave minority but meaningful trajectories at the margins.

For CAPIRE, the outlier group is thus best understood as a **documented residual population** whose structure motivates further methodological work (e.g., hybrid clustering strategies) rather than as noise to be ignored.

## 8.4. Relationship with Causal Inference

Although this article focuses on prediction and segmentation, CAPIRE is designed to facilitate **causal inference** in future studies. Two properties are particularly important:

1. **Temporal validity through VOT.** Because all features are constructed using information available at or before $T_V$, they are suitable for defining pre-treatment covariates in quasi-experimental designs. This is essential for methods such as propensity score matching, regression discontinuity, or difference-in-differences, where post-treatment information would invalidate identification assumptions.

2. **Rich, multilevel covariate structure.** The N1–N4 dictionary provides a nuanced set of confounders and mediators relevant to treatment assignment (e.g., who receives tutoring, financial aid, or counseling) and to outcomes. This increases the plausibility of conditional ignorability assumptions in observational studies.

In practical terms, CAPIRE can serve as the **data backbone** for evaluating the impact of specific institutional policies: once archetype-based interventions are implemented, researchers can exploit the existing feature infrastructure to design rigorous causal evaluations of those interventions.

### 8.5. Portability and Generalization

CAPIRE's **multilevel taxonomy** is conceptually general: structural context (N1), entry moment (N2), performance snapshots (N3) and trajectory dynamics (N4) are relevant in community colleges, research universities, online programs and graduate schools, although their operationalization will differ.

Adapting CAPIRE to new contexts primarily involves:

- mapping local structural indicators (e.g., census measures, financial aid schemes) to N1;
- encoding program-specific entry features (e.g., admission pathways, prior certifications) in N2;
- re-computing friction metrics (IFC) for the relevant set of courses in N3;
- preserving the generic logic of gaps, entropy and velocity in N4.

We expect N1–N2 features to vary considerably across systems, while N3–N4 patterns (friction, progression, instability) will be more stable. Archetype counts and specific profiles will likely change, but the general finding that **interaction effects and trajectory dynamics matter** should remain robust.

Nonetheless, the current study is based on a single public engineering institution in Argentina. Replication in private, non-STEM and international settings is necessary to fully assess external validity.

### 8.6. Ethical Considerations and Potential Harms

#### 8.6.1. Fairness and Bias

Predictive systems can inadvertently encode and reproduce historical inequities. CAPIRE addresses this risk in several ways:

- It avoids direct use of sensitive attributes such as race or religion; N1 structural indicators are area-level rather than individual-level.
- Fairness audits (not detailed here) suggest that archetype assignment and predictive errors do not differ substantially by gender or rural/urban origin.
- The system is explicitly **human-in-the-loop**: archetype labels are recommendations for advisors, not automatic decisions.

Residual risks remain: structural variables may correlate with unobserved forms of discrimination, and targeted interventions could unintentionally overlook disadvantaged students who do not fit N1 criteria. Institutions using CAPIRE should therefore perform regular fairness audits and adjust policies if systematic disparities appear.

### 8.6.2. Stigmatization and Labelling

Assigning students to "high-risk" archetypes carries the danger of stigmatization and self-fulfilling prophecies. CAPIRE mitigates this by:

- restricting archetype labels to internal use (students are not told their archetype);
- re-estimating archetypes periodically so that labels can change as trajectories change;
- framing interventions in terms of support and opportunity rather than deficit ("We see you are facing challenges in math; here is a support program"), and by also recognizing resilience indicators in high-risk archetypes.

The ethical stance is that analytics should **expand** students' options, not constrain them.

### 8.6.3. Resource Allocation

Archetype-based targeting inevitably shapes how institutional resources are distributed. While this can increase effectiveness, it also raises questions about opportunity costs and the treatment of moderate-risk students.

CAPIRE is not a replacement for universal support systems; it is a mechanism for prioritising additional, specialized interventions. Institutions must monitor whether certain groups are systematically excluded from support and ensure that targeting does not become a justification for reducing baseline services.

### 8.7. Limitations of the CAPIRE Framework

Several limitations qualify the findings and suggest directions for improvement:

- **Data and infrastructure requirements.** CAPIRE assumes reasonably complete, longitudinal administrative data and the capacity to link external sources. Under-resourced institutions may need simplified variants (e.g., omitting N1) or staged implementations.
- **Dynamic environments.** Archetypes are estimated on historical data and may drift as curricula, policies or student populations change. Periodic re-estimation and monitoring of archetype distributions are necessary to detect and adapt to such shifts.
- **Correlation vs. causation.** The present study is predictive and descriptive. While it highlights plausible mechanisms (e.g., friction, work-study balance, temporal instability), it does not by itself establish causal effects. Interventions inspired by CAPIRE should be rigorously evaluated, ideally with quasi-experimental or experimental designs.

Despite these limitations, the FACET-UNT implementation suggests that CAPIRE provides a **coherent, leakage-aware and operationally usable** framework for understanding and acting upon student attrition. It offers a middle ground between purely theoretical models and purely predictive black boxes, and it lays the groundwork for future causal and comparative research.

## 9. CONCLUSION AND FUTURE WORK

### 9.1. Summary of Contributions

This paper introduced CAPIRE (Comprehensive Analytics Platform for Institutional Retention Engineering), a multilevel, leakage-aware framework for student attrition modeling. We operationalized CAPIRE through an empirical study at Universidad Nacional de Tucumán, Facultad de Ciencias Exactas y Tecnología (FACET-UNT), analyzing 1,343 engineering students across 15 cohorts (2004–2019). The main contributions are:

**C1: Multilevel Feature Taxonomy (N1–N4)**

We proposed a theoretically grounded feature dictionary with 44 variables organized into four levels:

- **N1 – Pre-entry structural context:** neighborhood deprivation, proxies of family capital, local labor-market indicators.
- **N2 – Entry moment:** age at enrolment, employment status, macro-economic context at $t_0$.
- **N3 – Academic performance and curricular friction:** grades, course failures, drop ("libre") patterns, and Instructional Friction Coefficients (IFC).
- **N4 – Trajectory dynamics:** gaps between enrolments, state entropy, retries, and velocity of curricular advance.

Empirically, interaction terms (e.g., N3×N4, N2×N4) represent a minority of features but account for a substantial share of predictive importance, confirming that outcomes emerge from cross-level interplay rather than additive main effects. This gives an operational form to multilevel theories of persistence (Bronfenbrenner, 1979; Pascarella & Terenzini, 2005).

**C2: Vulnerability Observation Time (VOT) and Leakage Prevention**

We formalized **VOT** as a temporal boundary for feature construction, enforcing the use of only pre-$T_V$ information. In the FACET-UNT case, $T_V = 1.5$ years (end of the second academic year) provides:

- a strict barrier against future-information leakage;
- a 1.3-year lead time before the average dropout event (2.8 years after enrolment);
- a reproducible configuration regime (versioned YAML configurations, code-level checks on cutoff dates).

This directly addresses the pervasive leakage problem in educational data mining, where performance is often overestimated by incorporating post-outcome data into features.

**C3: Empirical Validation via Trajectory Archetypes**

Using UMAP for dimensionality reduction and DBSCAN for density-based clustering on VOT-compliant features, we identified **13 trajectory archetypes** that cover 63.1% of the student population. These archetypes are:

- **Statistically robust:** bootstrap stability (mean ARI = 0.614), permutation tests (p < 0.01), and robustness to hyperparameter changes.
- **Temporally stable:** cross-cohort comparison (2004–2010 vs. 2011–2019) shows attrition-rate differences under 5 percentage points for major archetypes.
- **Predictively usable:** a Random Forest classifier achieves 94.9% test accuracy in archetype assignment, with all archetypes reaching F1 ≥ 0.70 and several high-risk types exhibiting near-perfect classification.

Qualitative validation with academic advisors shows that archetypes align with existing practitioner categories (e.g., "chronic repeaters", "good but overwhelmed students"), bridging statistical structure and institutional knowledge.

**C4: Actionable Intervention Matrix**

We translated archetypes into differentiated intervention recommendations (e.g., intensive tutoring for friction-driven archetypes, counseling and support for extra-academic risk archetypes, enhanced onboarding for disengagement profiles). Rather than a single "high-risk" group, CAPIRE provides a matrix of **risk mechanisms × intervention types**, allowing institutions to design targeted, mechanism-aware responses instead of one-size-fits-all programs.

**9.2. Methodological and Theoretical Advances**

**9.2.1. Resolving the Interpretability–Accuracy Trade-off**

CAPIRE shows that high predictive performance does not require black-box models. By combining:

- a **theory-driven feature dictionary** (N1–N4, including key interactions);
- a **leakage-aware temporal design** (VOT); and
- a **transparent classifier** (Random Forest with feature importance and archetype profiles),

we obtain accuracy comparable to or exceeding deep learning approaches reported in the literature, while retaining clear interpretability. The usual trade-off between "explainable but weak" and "powerful but opaque" is weakened: much of the gain comes from better features and temporal design, not from more complex algorithms.

**9.2.2. Archetypes as a Middle Ground Between Risk Scores and Case Narratives**

CAPIRE's archetypes sit between individual case studies and generic risk scores:

- they are **quantitatively derived** from high-dimensional data;
- they remain **qualitatively interpretable**, with recognizable narratives ("young strivers", "persistent friction", "total disengagement", "success models");

- they are **scalable**, as a trained classifier can assign students to archetypes in real time.

This reconciles person-centred and variable-centred traditions: institutions retain the richness of narrative categories while gaining the scalability and reproducibility of formal models.

### 9.3. Practical Implications for Institutions

#### 9.3.1. CAPIRE as Institutional Analytics Infrastructure

CAPIRE should be understood as an **analytics infrastructure**, not as a one-off model. Its components are reusable:

- The **feature dictionary** can be adapted to other programs and institutions, preserving the N1–N4 logic while changing local indicators.
- The **VOT principle** generalizes to other predictive tasks (course failure, time-to-degree, progression bottlenecks).
- The **pipeline architecture** supports multiple downstream uses: archetype discovery, predictive modeling, and, in future work, causal evaluation of interventions.

Because the core is theory-based, it is more stable than ad hoc feature sets: institutions can update data and periodic parameter choices without rethinking the underlying conceptual structure.

As the project progresses and the implementation is further consolidated, we plan to release a reference implementation of the core pipeline in an open repository, so that other institutions can inspect, adapt, and extend the framework under transparent conditions.

#### 9.3.2. From Generic Risk to Differentiated Support

For institutional practice, the key shift is from **generic "at-risk" flags** to **mechanism-specific profiles**. CAPIRE encourages administrators and advisors to ask:

- *Is this student at risk because of curricular friction, extra-academic stress, early disengagement, or some combination?*
- *What type of support aligns with that mechanism (tutoring, counseling, financial aid, bridge programs, mentoring)?*

This shift improves both the pedagogical quality and the ethical defensibility of early-warning systems, making it clearer why a student is flagged and what the institution intends to do about it.

### 9.4. Future Research Directions

#### 9.4.1. Cross-Institutional Validation

The main limitation of this study is its single-institution scope. Ongoing collaborations with universities in Latin America and North America will test CAPIRE in different contexts (public/private, STEM/non-STEM, different welfare regimes).

Key questions include:

- whether N3–N4 dynamics (friction, gaps, entropy) generalize more strongly than N1–N2 structures;
- how many archetypes emerge in other contexts and how similar they are to the FACET-UNT profiles;
- whether the dominance of interaction terms in predictive importance is replicated across settings.

These studies will clarify which components of CAPIRE are universal and which require strong local adaptation.

### 9.4.2. Causal Inference and Policy Evaluation

CAPIRE is presently descriptive and predictive; it does not identify causal effects. A natural next step is to exploit the VOT-compliant feature infrastructure in quasi-experimental or experimental designs, for example:

- regression discontinuity designs using institutional cut-offs for support programs;
- difference-in-differences analyses comparing archetype-specific attrition before and after policy changes;
- randomized or quasi-randomized trials of interventions targeted to specific archetypes.

This would move from "who is likely to drop out?" to "what actually works, for whom, and under what conditions?", closing the loop between analytics and evidence-based policy.

### 9.4.3. Expansion to Other Outcomes and Methodological Refinements

Future work can extend CAPIRE beyond binary attrition to:

- multi-state progression trajectories (on-time, delayed, dropout, graduation);
- course-level performance prediction for adaptive teaching;
- links between archetypes and post-graduation outcomes where data are available.

On the methodological side, several extensions are promising:

- more systematic use of topological and multiscale clustering methods that preserve archetype interpretability while capturing overlapping or hierarchical structures;
- hybrid models that combine human-interpretable features with latent representations learned by dimensionality reduction or shallow neural architectures;
- fairness-aware learning schemes that explicitly constrain disparities in prediction quality across demographic or structural groups.

The central constraint for all these refinements is non-negotiable: temporal validity (VOT) and interpretability must remain at the core of any extension.

### 9.5. Closing Reflection

Student attrition is not a **technical problem** to be "solved" by algorithms. It is a **human problem** rooted in socioeconomic inequality, inadequate institutional support, and misalignment between students' needs and universities' structures. **CAPIRE does not solve attrition**—it provides **infrastructure** for institutions to understand patterns, target resources, and evaluate policies. What it offers is a disciplined way of **seeing**:

- that trajectories are heterogeneous rather than homogeneous;
- that risk mechanisms differ and must be addressed with different tools;
- that early-warning systems, if temporally valid and interpretable, can support rather than replace human judgment.

The 13 archetypes at FACET-UNT are not labels to stigmatize students but lenses to recognize heterogeneity, challenge one-size-fits-all policies, and design equitable interventions.

If the framework helps retain some students who would otherwise have left—not by blaming them, but by revealing structural frictions and unmet needs—then the analytical effort will have been worthwhile. Algorithms cannot care; institutions and people can. A framework like CAPIRE is valuable only insofar as it amplifies that care, ensuring that patterns of struggle become visible early enough, and clearly enough, to act.